\documentclass[a4paper,11pt]{article}
\usepackage{jcappub} 

\title{\boldmath Constraining Palatini gravity with GR-independent equations of state for neutron stars}

\author[a]{Eva Lope-Oter,}
\affiliation[a]{Departamento de F\'isica Te\'orica \& IPARCOS, Universidad Complutense de Madrid, E-28040, 
Madrid, Spain}
\author[a]{Aneta Wojnar}

\emailAdd{awojnar@ucm.es}

\abstract{We demonstrate how to construct GR-independent equations of state for a neutron star from the information available in the literature. We emphasize the importance of using theory-based principles instead of relying solely on astrophysical observables and General Relativity. We propose a set of equations of state based on first microscopic principles, including chiral perturbation theory and perturbation theory in quantum chromodynamics. Interpolation methods are employed with assumptions on the thermodynamic stability and causality in the intermediate region. These equations of state are then used to constrain quadratic Palatini $f(\mathcal R)$ gravity, indicating that its free parameter  can at most lie within the range around  $-6.47 \lesssim \beta \lesssim 1.99$ km$^2$. Additionally, we briefly discuss the problem of phase transitions and twin stars.}

\begin{document}
\maketitle
\flushbottom


\section{Introduction}

General Relativity (GR) has proven to be highly predictive across a wide range of situations, from accurately describing the Solar System to explaining the behavior of black holes. GR has also been confirmed by the detection of gravitational waves from the merger of two black holes \cite{abbott2016observation} and, soon after, from the merger of two neutron stars (NSs) \cite{abbott2017gw170817}. However, GR has its limitations in explaining various cosmological and astrophysical phenomena. For example, it fails to account for the presence of dark matter in astrophysics \cite{rubin1980rotational}, the accelerated expansion of the Universe (the dark energy problem) \cite{huterer1999prospects}, or the early inflation in cosmology, which has left many unanswered questions \cite{copeland2006dynamics,nojiri2007introduction,nojiri2017modified,nojiri2011unified,capozziello2008extended,CANTATA:2021ktz}.

The maximum mass of a neutron star (NS) is a crucial issue, and it is closely related to both the equation of state (EoS) and the adopted theory of gravity. As a result, researchers must address the so-called mass-radius (M-R) degeneracy. A specific point on the M-R diagram can correspond to a particular EoS in General Relativity (GR) or to a different EoS within the framework of Extended Theories of Gravity (ETG). For more information on this topic, refer to \cite{Olmo:2019flu} and the references therein.

The initial theoretical mass limit for neutron stars was estimated to be around $M \approx 0.6 M_\odot$ using the free neutron gas equation of state (EoS) and the Tolman-Oppenheimer-Volkoff (TOV) equation within the framework of GR \cite{1939PhRv...55..364T}. However, subsequent research has proposed numerous other EoSs, resulting in a significant increase in the maximum mass limit to approximately three times the initial value \cite{Ozel:2016oaf}. Nevertheless, recent observations have raised questions about the physics underlying equations of state. The measurement of a neutron star with a mass of approximately $M \approx 2.4 M_\odot$ \cite{Romani:2022jhd} has challenged current theoretical models. Additionally, the maximum masses of observed compact objects may exceed theoretical predictions, as suggested by the data from the GW190814 event \cite{Abbott:2020khf}, which implies that the secondary component should have a gravitational mass greater than 2.5 $M_\odot$, potentially indicating the existence of either a heavy neutron star or a light black hole.

The maximum allowed mass of a NS is crucial in understanding the evolution of stellar remnants and determining the final outcome of core-collapse supernovae and binary neutron star mergers. The maximum mass depends primarily on the EoS at densities higher than three times the nuclear saturation density $n_s = 0.16$ fm$^{-3}$, while the radii of canonical NS's with mass around $1.4M_\odot$ are largely determined by the EoS at densities $n<3n_s$ \cite{Lattimer:2004pg}. Maintaining a slope $c_s^2=1$ up to 5 times the nuclear saturation density requires a form of strongly interacting relativistic matter, which poses significant challenges for dense matter theory and Quantum Chromodynamics (QCD) \cite{McLerran:2018hbz}.

Furthermore, it has been hypothesized that additional degrees of freedom can appear in the core at high densities. However, the presence of these additional degrees of freedom softens the equation of state (EoS), making it difficult to achieve maximum masses higher than $2M_\odot$. This gives rise to the hyperon puzzle, which refers to the challenge of reconciling the measured masses of neutron stars (NSs) with the presence of hyperons or any other degree of freedom in their interior \cite{Blaschke:2018mqw,Baym:2017whm,Chatterjee:2015pua,Annala:2019puf,McLerran:2018hbz,Leonhardt:2019fua}. These challenges motivate the exploration of alternative frameworks in the physics of compact objects, such as the previously mentioned Extended Theories of Gravity (ETG) \cite{Joyce:2016vqv,Olmo:2019flu}.

Among the various theories beyond GR, this work specifically investigates Palatini $f(\mathcal{R})$ gravity, which is one of the simplest metric-affine models of gravity \cite{gronwald1997metric,de2010f,baldazzi2022metric}. In order to compare GR with modified theories of gravity in the context of dense matter, it is crucial to have a priori knowledge of the equation of state (EoS) derived from nuclear and hadron physics. However, this EoS should not be subject to any astrophysical constraints because those already assume GR, and therefore, cannot be utilized to constrain theories beyond GR.

In this study, we present several equations of state (EoSs) that are solely constrained by inputs from hadron physics and fundamental principles, without incorporating observations of neutron stars (NSs). These EoSs are constructed by incorporating state-of-the-art results from Chiral Effective Field Theories ($\chi$EFT) for the low-density regime, while for the very high-density regime, we employ methods and outcomes from perturbative Quantum Chromodynamics (pQCD). To bridge the intermediate density region, we employ interpolation techniques based on first principles, including thermodynamic stability, consistency, and causality. These principles serve as the sole restrictions for obtaining the EoSs \cite{LopeOter:2019pcq,Komoltsev:2021jzg}. As our EoSs are less constrained compared to those commonly used in the literature, they are more suitable for use in ETG and serve as valuable tools for testing these proposals.

The paper is structured as follows: In Section \ref{MGvGR}, we provide a brief overview of the fundamental concepts related to equations of state (EoSs) and discuss the methodology employed in this study. Section \ref{palsec} presents a concise description of Palatini $f(\mathcal R)$ gravity, which enables us to apply it to both models of gravity discussed in Section \ref{MGresults}. Finally, in the concluding section, we summarize our findings and present our conclusions.

Throughout the paper, we adopt the convention of $(-,+,+,+)$ for the signature of the metric and use the natural units $G=c=1$, leading to $\kappa^2=8\pi$.


\section{Equations of state}\label{MGvGR}

The majority of EoSs describing the matter inside neutron stars are derived based on astrophysical observables and within the framework of GR \cite{Capano:2019eae,Most:2018hfd, Tews:2018iwm,Annala:2019puf, Dietrich:2020efo, Essick:2021kjb,Raaijmakers:2021uju, Annala:2021gom,Huth:2021bsp,Altiparmak:2022bke,Gorda:2022jvk,Gorda:2022lsk}. Consequently, it is inconsistent to utilize these EoSs in the context of ETG. Moreover, it has been demonstrated that the choice of the gravitational theory does impact the derived equations of state \cite{kulikov1995low,kim2014physics,hossain2021equation,hossain2021higher,wojnar2023fermi,pachol2023fermi}. Therefore, it is crucial to deploy EoSs that do not assume any dependence on the aforementioned features.

Specifically assuming GR, many equations of state have been ruled out in order to satisfy the astrophysical constraints. Examples of this selection process include soft equations of state \cite{radice2018gw170817} or EoSs with multiple phase transitions, such as those involving hyperons \cite{Katayama:2012ge,Lonardoni:2013gta,Petschauer:2015nea,Burgio:2021vgk} or quark matter \cite{Weissenborn:2011qu, Annala:2019puf}. If GR is modified, these analysis become void.

In what follows, we will construct equations of state (EoSs) that are independent of the specific theory of gravity by relying solely on theoretical first principles. Importantly, we will not utilize any observational constraints, which is crucial as these measurements are often performed assuming General Relativity (GR) \footnote{For example, NS-NS merger simulations assume GR, while gravitational wave detectors are not sensitive to modes that appear in Extended Theories of Gravity (ETG) but do not exist in GR \cite{Mendes:2018qwo}.}. Additionally, we will not employ other phenomenological model-dependent EoSs, as they lack uncertainty bands \cite{Akmal:1998cf,Chamel:2008aa,Baldo:2012hw,Douchin:2001sv,Typel:2009sy,Hempel:2009mc,Fattoyev:2010mx}. The objective is to construct EoSs that are as model-independent as possible, devoid of astrophysical (and GR) inputs.

Consequently, our EoSs are less constrained compared to those found in the existing literature. This makes them more reliable for utilization within the framework of ETG and for testing gravitational proposals or Beyond Standard Model physics. Consequently, we can make a meaningful comparison between GR and a specific model of gravity, while also conducting a comprehensive analysis of the impact of ETG on the mass-radius diagrams, stability, and other properties of neutron stars.

Let us first introduce some basic aspects related to neutron stars and the equations of state.
To describe the structure of NS, the EoS of neutron matter at zero temperature ($T = 0$) is a crucial component. It relates the relationship between the pressure and the energy density ($\varepsilon$) of the matter, i.e., $p(\varepsilon)$. Additionally, an important quantity associated with the EoS is the speed of sound, defined as $c_s^2 = dp/d\varepsilon$, which characterizes the propagation of perturbations in the matter and depends on the specific EoS.

A neutron star is typically divided into three main regions: the atmosphere, the crust, and the core. The central density of a neutron star is often expressed in terms of the baryon number density $n$ (nucleons per fm$^3$) related to the saturation density number $n_s$. The saturation density is defined as the density of symmetric nuclear matter, where the binding energy per nucleon is at its minimum. The value of the saturation density is approximately $n_s=0.16$ fm$^{-3}$, which corresponds to approximately $2.7 \times 10^{14}$ g cm$^{-3}$. The central density of a neutron star can reach values that are 5 to 10 times higher than the saturation density.

The crust of a neutron star is a seemingly solid layer that is relatively thin, only a few hundred meters in thickness. It is the core of the star that contributes the most to its radius and mass. The crust can be divided into two main regions: outer and inner.
 The outer crust consists of a solid lattice structure of heavy nuclei, predominantly around the mass number of iron. These nuclei exist in a Coulomb lattice and are in $\beta$-equilibrium, which means that neutron decay is balanced by the inverse process or electron capture. Along with the heavy nuclei, there is also an electron gas present in this region.
Moving towards the center of the neutron star, the density increases, and the nuclei in the crust progressively become more  neutron-rich. Eventually, at a certain point known as the neutron-drip point, the density becomes high enough that neutrons start to leak out from the nuclei, and the inner crust begins. There, the crustal matter consists of a mixture of free neutrons, a smaller fraction of free protons, and electrons.
At even higher densities, it is believed that the crust dissolves into a sea of predominantly free neutrons, along with a smaller fraction of free protons and electrons \cite{Blaschke:2018mqw}.

At approximately half of the saturation density ($n \approx 0.08$ fm$^{-3}$), the nuclei in the neutron star crust completely dissolve, and the core region begins. The core of a neutron star is much less understood than the crust and is also divided into outer and inner parts.
The outer core of a neutron star is believed to be  composed of uniform matter, primarily neutrons with a small fraction of protons and electrons. This matter is in $\beta$-equilibrium, balancing the chemical potentials of neutrons ($\mu_n$), protons ($\mu_p$), and electrons ($\mu_e$) without the presence of neutrinos and muons, $\mu_n - \mu_p = \mu_e$. Since the proton fraction in the uniform nucleonic matter is small, approximately 5\%, neutron-star matter can be effectively described as pure neutron matter.
At even higher densities, above approximately twice the saturation density ($n \approx 2n_s$), the composition of matter in the core becomes largely unknown.

The internal constitution of the outer crust is well known, and the EoS of this region is determined from experimental nuclear physics \cite{Chamel:2020hni}. However, the inner crust  cannot be reproduced in terrestrial laboratories, and we must rely on theoretical models.
Regarding the core, there are various models that describe the relation between the total pressure and energy density for either pure neutron matter or $\beta$-equilibrium neutron star matter. These models can be classified into two main categories: microscopic many-body approaches and phenomenological models.

The phenomenological approaches consist of specific physical models, such as relativistic field models with $\sigma$, $\omega$, and $\rho$ meson exchanges, hyperons, or models with explicit quarks. Therefore, the resulting equations of state (EoSs) within these approaches are generally dependent on the chosen model. Examples of these EoSs are APR \cite{Akmal:1998cf}, BSK \cite{Chamel:2008aa}, BCPM \cite{Baldo:2012hw}, Sly \cite{Douchin:2001sv}, DD2 \cite{Typel:2009sy}, HLPS \cite{Hempel:2009mc}, and IUF \cite{Fattoyev:2010mx}.

The ab-initio approach is a microscopic understanding of the structure of atomic nuclei and dense matter based on the properties of the fundamental degrees of freedom and their interactions. In this approach, $\chi$EFT has emerged as a powerful tool for performing microscopic calculations of nuclear matter properties up to approximately 2$n_s$. $\chi$EFT is an effective theory of QCD that describes strong interactions using nucleon and pion degrees of freedom. The range of validity of the $\chi$EFT expansion is determined by the breakdown scale $\Lambda_B$ (also known as the cutoff), which is estimated to be around 500-600 MeV/c. Furthermore, this theory provides an ordering scheme characterized by an expansion of the nuclear potential in terms of momenta divided by the breakdown scale $\Lambda_B$, which determines the order of the expansion. The main advantage of $\chi$EFT compared to phenomenological approaches (regardless of the specific quality of the description of particular nuclear datasets) is that theoretical uncertainties can be quantified by analyzing the convergence of the $\chi$EFT expansion order by order. However, the applicability of this theory is limited to baryon numbers $n$ of about 2 times the nuclear saturation density $n_s \approx$ 0.16 fm$^{-3}$.
 
\begin{figure}
     \centering
     \includegraphics[width=6.5in]{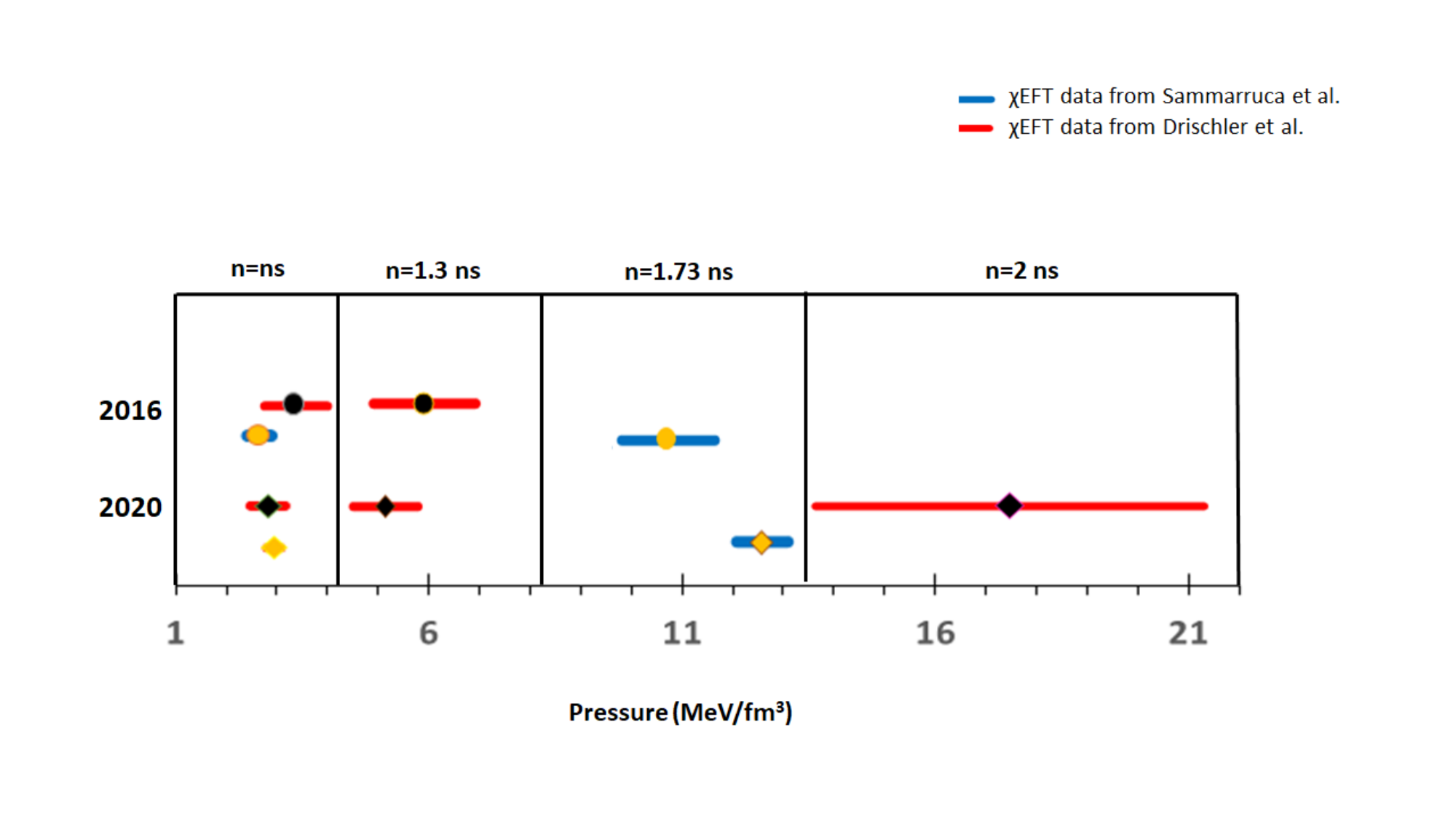}
 \caption{[Color online] The computational improvement in the last few years (2016 - 2020), has allowed the uncertainties to decrease. The pressure is plotted at the matching point between $\chi$EFT and our interpolation, with data provided by Sammarruca et al. \cite{Sammarruca:2016ajl,Sammarruca:2018whh} (blue lines) and by Drischler et al. \cite{Drischler:2016djf,Drischler:2020hwi} (red lines). The red/yellow diamond (Drischler/Sammmarruca) stands for 2020 and the circle for 2016 (in the same colors).}
     \label{Errorbar}
 \end{figure}

In spite of that, there has been significant progress in recent years in quantifying uncertainties. A new framework has been developed \cite{Melendez:2019izc} to quantify correlated truncation errors in zero-temperature EFT calculations. This framework is based on recent order-by-order many-body perturbation theory calculations in pure neutron matter and symmetric matter with chiral two- and three-nucleon interactions up to\footnote{The abbreviation N$^3$LO refers to next-to-next-to-next-to-leading order, corresponding to the fourth order.} N$^3$LO.
Using this framework, reliable calculations have been performed up to approximately 2$n_s$ \cite{Drischler:2020hwi,Drischler:2020yad}. Therefore, the theoretical uncertainty (in terms of pressure) is about $\pm 25\%$ at $n=2n_s$, but only about $10\%$ at $n=n_s$ for pure neutron matter at N$^3$LO and a cutoff $\Lambda=500$ MeV \cite{Drischler:2020fvz,Melendez:2019izc}. Figure \ref{Errorbar} demonstrates the significant decrease in the uncertainty of the data provided by \cite{Sammarruca:2016ajl,Sammarruca:2018whh,Drischler:2016djf,Drischler:2020hwi}.

On the other hand, at very high densities ($n>40n_s$), which greatly exceed the densities in the NS core, the EoS of dense matter is known in the pQCD regime. The pQCD calculations have yielded robust results with their inherent uncertainties \cite{Gorda:2021kme}. However, reaching this pQCD regime is unlikely within a neutron star, as it indicates a limit to the pressure, energy density, and baryon chemical potential at the pQCD matching point. This provides a constraint on the EoS, creating a boundary through which the EoS must pass.

Nevertheless, neutron stars are expected to reach densities higher than 2$n_s$ (in the range of $5-10 \, n_s$). While the EoS of dense matter is known at very low or very high densities, it is not well-known in the energy density range required to describe the internal properties of a neutron star. Consequently, between these two extreme regions (the chiral domain and the pQCD regime), there exists a region where the equation of state needs to be approximated using various interpolation techniques.

Therefore, we will construct the equation of state (EoS) of neutron stars by combining different approaches and matching them in various density regions. For the low-density regime, we will utilize $\chi$EFT calculations in pure neutron matter and $\beta$-equilibrated matter at N$^3$LO, as presented in \cite{Drischler:2020hwi,Drischler:2020yad}. These calculations cover the density range of $n = 0.05 - 0.34$ fm$^{-3}$, which includes densities of the inner crust to the border with the inner core of neutron stars. 
At very high densities ($n>40n_s$), where the pQCD regime is applicable, we will incorporate the partial N$^3$LO results from \cite{Gorda:2021kme}. Notably, these approaches are advantageous as they are based solely on QCD symmetries and provide uncertainty bands for their predictions.

However, in the intermediate density region between the chiral domain and the pQCD regime (corresponding to the neutron star's inner core), we will employ interpolation techniques to approximate the equation of state. Our goal is to construct a set of $\beta$-equilibrated neutron star matter EoS based on first principles, specifically $\chi$EFT, and perturbation theory in QCD. This interpolation will allow us to bridge the gap between the known EoS in the low and high-density regions.

\subsection{Constructing EoS's}

Let us now explore the construction of the EoS based on first principles, specifically on the foundation of QCD, as mentioned earlier. It is worth noting that EoS can be categorized into two subclasses: stiff and soft. Stiff EoS refers to those with large slopes near $c_s^2\lesssim 1$  and they cause the radius of the neutron star to increase (or remain roughly constant in certain regions of the mass-radius diagram) as the mass increases. In other words, the pressure inside the compact object is sufficiently high to counteract gravitational attraction. On the contrary, soft EoS (with small slopes) leads to a decrease in the star's radius as the mass increases.

As already discussed, we are dealing with three regimes of density:
\begin{itemize}
    \item \textbf{Low ($\chi$EFT regime)}\\
At low densities $0.05 \leq n \leq 0.34$ fm$^{-3}$, we utilize the data from \cite{Drischler:2020hwi,Drischler:2020yad}. Since the $\chi$EFT EoS are provided in terms of $n$ (number density) and the binding energy per nucleon $E/A$, we extract the EoS (pressure $P$ as a function of the energy density $\varepsilon$) from their data using the following relations:
\begin{equation}\label{eq:densityenergy}
\varepsilon= n\left(M_N+\frac{E}{A}\right)
\end{equation}
\begin{equation}\label{eq:firstlaw}
P= n^2\frac{d(E/A) }{d n},
\end{equation}
where \eqref{eq:densityenergy} represents the expression for the energy density, including the rest mass of the particles, and \eqref{eq:firstlaw} corresponds to the first law of thermodynamics at $T=0$. By using cutoffs at $450$ and $500$ MeV from \cite{Drischler:2020hwi,Drischler:2020yad}, we obtain an uncertainty band for this region.
\item \textbf{Very high (pQCD regime)}\\
For densities $n \geq 40 n_s$, we consider as a matching point with the pQCD region a specific baryon chemical potential value $\mu_B \approx 2.6$ GeV, within an uncertainty band between different values of the scale parameter $X$ ranging from $1$ to $4$\footnote{The scale parameter $X\equiv 3\bar{\Lambda}/\mu_B$, where $\bar{\Lambda}$ is the renormalization scale.}. Similar to the $\chi$EFT region, we utilize the energy density, pressure, and number density values from \cite{Gorda:2021znl} for these two values of $X$ as bounds to perform the interpolation in this high-density regime.
  \item \textbf{Interpolation regime} \\
In the intermediate region between the chiral domain and the pQCD regime, we establish the maximum allowable region based on the conditions of causality ($c^2_s\leq 1$) and monotonic behavior ($c^2_s\geq 0$). Within this region, we construct a grid of candidate points ($\varepsilon, P$) representing potential equations of state (EoS), following the interpolation procedure developed in \cite{LopeOter:2019pcq}. The EoS can be generated in two different ways: randomly \cite{LopeOter:2019pcq} or by controlling the slope\footnote{Referring to the sound speed $c_s^2$.}. The latter approach involves extending the growth rate of the slope from the chiral region into the intermediate zone, as far as the grid permits \cite{Evaphd}. To accomplish this, we construct a 100-point EoS using a 1000 $\times$ 1000-point grid to achieve better control over the slope. The specific details of this procedure, employed in the subsequent parts of this paper, have been discussed in \cite{LopeOter:2019pcq,Evaphd}.
\end{itemize}

\begin{figure}
    \centering
    \includegraphics[width=5.4in]{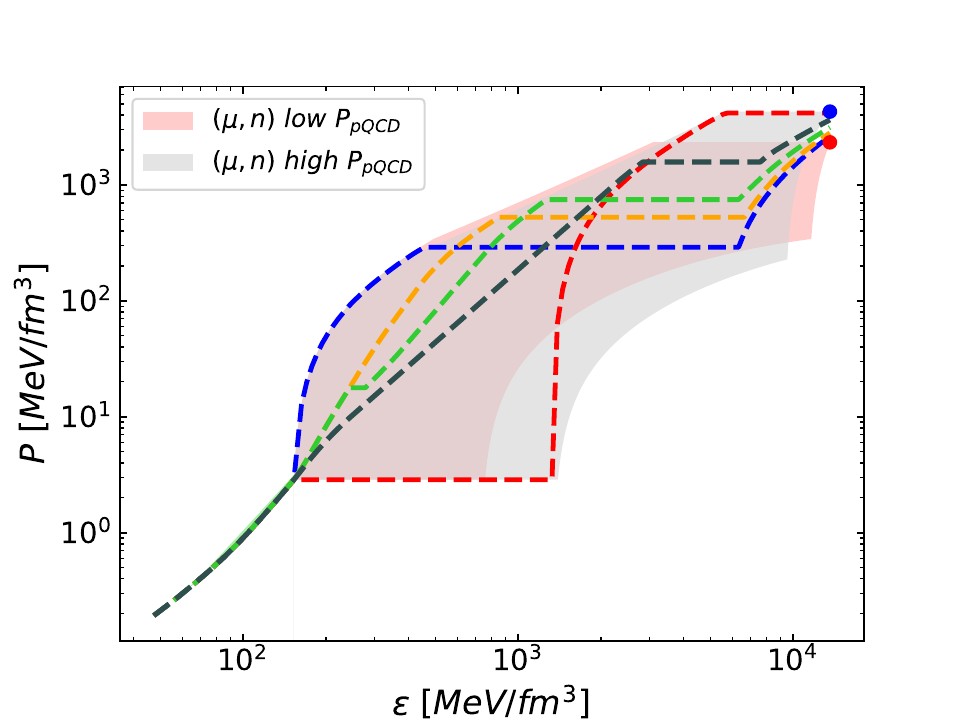}
    \includegraphics[width=5.4in]{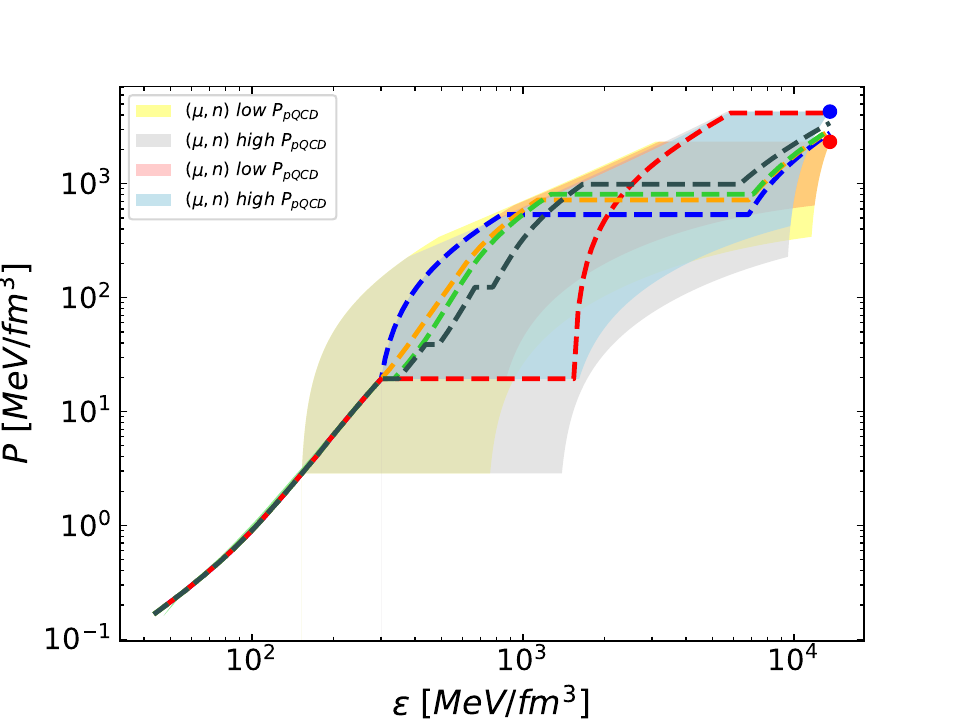}
    \caption{[Color online] Example set of EoSs with $\chi$EFT data from \cite{Drischler:2020yad} and pQCD limits from \cite{Gorda:2021znl} at the matching point $n=n_s$ (top), $n= 2n_s$  (bottom). The foreground (grey) and background (red) areas correspond to the constraints resulting from simultaneously imposing the limit for $n$ and $\mu_B$  at the high-$\varepsilon$ pQCD and the low-$\varepsilon$ pQCD  points, respectively.  The blue line always represents the stiffest EoS, while the red line represents the softest EoS (inside a neutron star) allowed by the pQCD limits. The two bands obtained by interpolating are shown at the bottom: at n=n$_s$ (yellow and gray) and at n=2n$_s$ (pale red and light blue).} 
    \label{fig:EoSinterpol2}
\end{figure}

 In what follows, we provide a more detailed explanation of the procedure employed in the interpolation zone. Initially, we construct equations of state (EoS) based on the candidate points of the grid, taking into account causality and monotonicity between energy density and pressure. However, it is also essential to ensure thermodynamic consistency, which we achieve by utilizing the following discrete equations:
\begin{eqnarray}
n_i &=& \frac{\varepsilon_i}{M_N+(E/A)i}\label{eq:discrenumberdensity}\\
P_i &=& n^2_i \frac{(E/A){i+1}-(E/A){i-1}}{n{i+1}-n_{i-1}}
\label{eq:discrefirstlaw}\\
\mu_{Bi} &=& \frac{\varepsilon_i+ P_i}{n_i},
\label{eq:Euler}
\end{eqnarray}
Here, equations \eqref{eq:discrenumberdensity} and \eqref{eq:discrefirstlaw} correspond to equations \eqref{eq:densityenergy} and \eqref{eq:firstlaw}, respectively, expressed for the discrete points derived from the grid. Equation \eqref{eq:Euler} represents the Euler equation, which ensures the aforementioned thermodynamic consistency. We impose constraints on the limiting values of ($\varepsilon, P, \mu_B, n$) obtained from the pQCD regime, which suggests a long first-order phase transition in order to achieve the desired baryon chemical potential $\mu_B=2.6$ GeV. Except for the very stiff EoSs (e.g., the Blue, Orange and Green EoSs of both the upper and lower part of Fig.\ref{fig:EoSinterpol2}), such phase transitions may happen at densities exceeding those of neutron stars, so that it is not a contradiction \cite{Brandes:2023hma}. Remarkably, these constraints derived from thermodynamic consistency lead to a softening of the EoS.

The obtained equations of state (EoS) are derived solely from hadron physics. Notably, in addition to providing information on energy density $\varepsilon$ and pressure $P$, our approach also includes details such as baryon density number $n$, binding energy per nucleon $E/A$, sound speed $c_s^2$ (representing the slope), and baryon chemical potential $\mu_B$. This comprehensive characterization sets our work apart from previous studies, such as \cite{Annala:2019puf, Altiparmak:2022bke}.

To account for the increasing chiral uncertainties with density number, ranging from $10\%$ at $1n_s$ to $25\%$ at $2n_s$, we perform interpolations that start at different matching points to the EoS obtained from $\chi$EFT. In this study, we present results from two interpolations: one taking off at $1n_s$ and the other at $2n_s$. We will also briefly discuss an interpolation at $1.5n_s$ We discuss these interpolations in detail in Section \ref{MGresults}.

From these procedures, we have selected five representative examples from our EoSs: the stiffest EoS, the softest EoS, and three intermediate ones.

\begin{figure}    
 \includegraphics{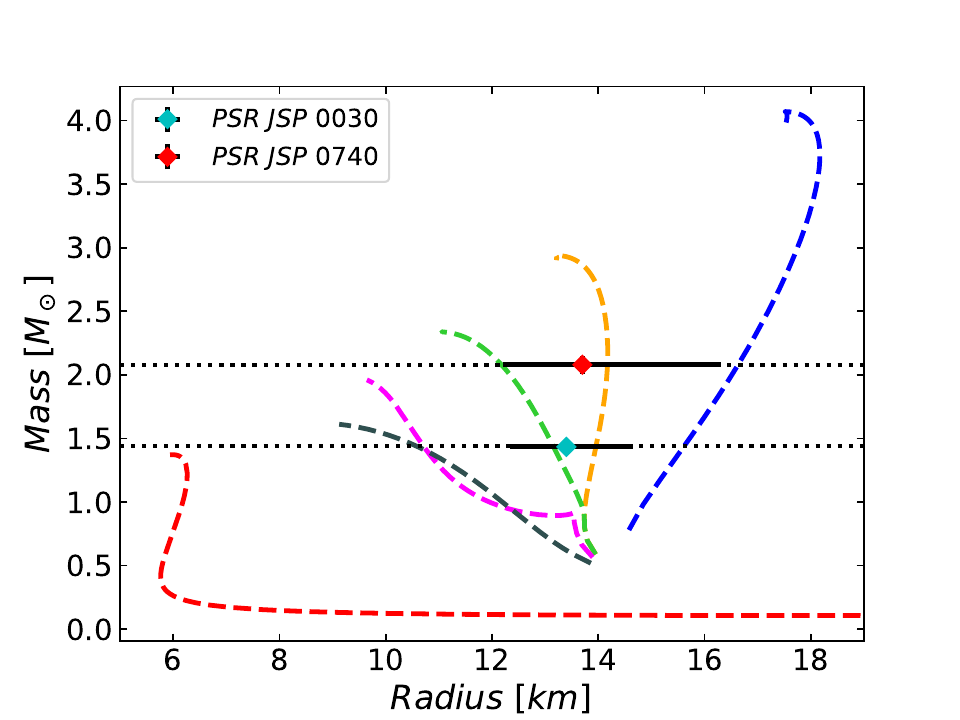}
    \caption{[Color online]Mass-Radius diagrams for EoSs shown at the top of Fig. \ref{fig:EoSinterpol2} for GR, depicting the mass and radius values of two recently observed pulsars. The horizontal dotted lines represent masses of 1.4 $M_\odot$ and $2.1 M_\odot$, respectively.}
    \label{fig:MvRinterpol2}
\end{figure}

\begin{table}
    \centering
    \begin{tabular}{ccccc}
    \hline
      EoS & $M_{max}/M_{\odot}$ & R$_{M_{max}}$ & \ $\varepsilon_c$ (MeV/fm$^3$)   \\ \hline \hline
        Blue & 4.07 & 17.52 & 455 \\
        Orange & 2.94 & 13.19 & 850\\
        Green & 2.34 & 11.06 & 1120\\ 
        Grey & 1.61 & 9.12 & 1958 \\
        Red & 1.37 & 5.99 & 4031 
    \end{tabular}
    \caption{Values of the maximum mass and corresponding radius and central energy density in GR for the EoSs of Fig. \ref{fig:EoSinterpol2}, for the interpolation at $n=n_s$. }
    \label{tab:GR1} 
\end{table}

\begin{table}
    \centering
    \begin{tabular}{ccccc}
    \hline
      EoS & $M_{max}/M_{\odot}$ & R$_{M_{max}}$ & \ $\varepsilon_c$ (MeV/fm$^3$)  \\ \hline \hline
        Blue & 2.99 & 13.1 & 834 \\
        Orange & 2.52 & 11.4 & 1135  \\
       Green & 2.37 & 10.09 & 1222\\
       Grey & 1.98 & 9.8 & 1432 \\
        Red & 1.29 & 5.94 & 4562 
    \end{tabular}
    \caption{Values of the maximum mass and corresponding radius and central energy density in GR for the EoSs of Fig. \ref{fig:EoSinterpol2}, interpolating at $n=2n_s$.} 
    \label{tab:GR2}
\end{table}

The stiffest equation of state (Blue EoS) used in this paper exhibits the maximum slope from the matching point of $\chi$EFT, as far as possible, to a point with a baryonic chemical potential $\mu_B \approx 1.9$ GeV. This bending value of $\mu_B$ in the interpolation zone allows us to reach $\mu_B = 2.6$ GeV in the pQCD regime \cite{Komoltsev:2021jzg}. In other words, starting from $\mu_B = 1.9$ GeV in the interpolation region, a first-order phase transition occurs until a point where the EoS transitions to pQCD with a conformal slope ($c_s^2 = 1/3$).

On the other hand, the softest equation of state (Red EoS) has the minimum slope ($c_s^2 = 0$) from the matching point of $\chi$EFT to a point in the interpolation region, which is determined by the constraints arising from the highest pressure of pQCD. From this point onwards, the Red EoS increases with the conformal slope ($c_s^2 = 1/3$) until it reaches the pQCD regime.

The three intermediate EoSs include the stiffest intermediate (Orange EoS), intermediate (Green EoS), and the softest intermediate (Grey EoS). These EoSs inherit their slopes from the $\chi$EFT region, but in the interpolation region, each EoS exhibits a different rate of slope increase.

Fig. \ref{fig:EoSinterpol2} illustrates all the discussed EoSs at the matching points $n=n_2$ and $n=2n_s$. The corresponding mass-radius diagram within the framework of General Relativity (GR) is shown in Fig. \ref{fig:MvRinterpol2}. Additionally, Tables \ref{tab:GR1} and \ref{tab:GR2} provide information on the maximum masses and the corresponding radii obtained using these EoSs in the context of GR.

By applying these EoSs to the TOV equation derived from a given ETG and solving it using the Runge-Kutta method, one can obtain solutions in the form of mass-radius diagrams. To proceed further, let us briefly discuss a chosen gravity model and the resulting TOV and mass equations.


 \section{Modified gravity: a brief description of Palatini $f(\mathcal R)$ gravity}\label{palsec}

Palatini $f(\mathcal{R})$ gravity is one of the simplest cases of the so-called metric-affine models of gravity. This approach treats the metric $g$ and the connection $\hat\Gamma$ as independent variables. Therefore, to obtain the field equations, the action of this theory of gravity must be varied with respect to both variables. The action is given by:
\begin{equation}\label{Eq:f(R)}
S=\frac{1}{2 \kappa^2} \int d^4x \sqrt{-g} f(\mathcal{R}) + S_m[g_{\mu\nu},\psi_m].
\end{equation}
Here, $\sqrt{-g}$ is the determinant of the metric, $f(\mathcal{R})$ is a function of the curvature scalar $\mathcal{R}$, and $S_m$ is the action for matter fields $\psi_m$, which depends only on the metric $g_{\mu\nu}$. It is important to note that the matter action $S_m$ is independent of the connection $\hat\Gamma$. While it is possible to generalize the matter action to include a dependence on $\hat\Gamma$, such an extension is expected to be relevant in the context of fermionic interactions rather than large-scale classical fluids. Hence, we will not consider it in this work.

Furthermore, the curvature scalar $\mathcal{R}$ is constructed from the metric and the Palatini-Ricci curvature tensor $\mathcal{R}_{\mu\nu}(\Gamma)$. Specifically, we have $\mathcal{R} = g^{\mu\nu}\mathcal{R}_{\mu\nu}(\Gamma)$. It is important to note that $\mathcal{R}{\mu\nu}$ must be symmetric in order to avoid instabilities \cite{alfonso2017trivial,beltran2019ghosts,jimenez2020instabilities}.

If the functional $f$ is linear in $\mathcal{R}$, we recover GR. In the case of vacuum or a pure radiation-dominated spacetime, we are dealing with GR with a cosmological constant. However, considering $f$ as an arbitrary function of $\mathcal{R}$ leads to a different spacetime structure \cite{Allemandi:2004ca,Allemandi:2004wn,olmo2011palatini}. Importantly, for our case, these modifications have implications for stellar descriptions \cite{Olmo:2019flu}. Let us explore this topic in more detail.

The variation of Eq. \eqref{Eq:f(R)} with respect to the metric $g_{\mu\nu}$ alone yields the field equations:
\begin{equation}
f'(\mathcal{R})\mathcal{R}_{\mu\nu}-\frac{1}{2}f(\mathcal{R})g_{\mu\nu}=\kappa^2 T_{\mu\nu},\label{structural}
\end{equation}
where $T_{\mu\nu}$ is the energy-momentum tensor of the matter field, obtained in the standard way:
\begin{equation}\label{Eq: energytensor}
T_{\mu\nu}=-\frac{2}{\sqrt{-g}}\frac{\delta S_m}{\delta g_{\mu\nu}}.
\end{equation}
In our case, we assume the matter content to be a perfect fluid. In the above equations, the primes denote derivatives with respect to the argument of the function: $f'(\mathcal{R})=df(\mathcal{R})/d\mathcal{R}$. It is evident that for linear $f(\mathcal{R})$, the equations \eqref{structural} reduce to those of GR.

On the other hand, the variation with respect to the independent connection $\hat\Gamma$ can be expressed as \cite{Wojnar:2017tmy}:
\begin{equation}
\hat\nabla_\beta(\sqrt{-g}f'(\mathcal{R})g^{\mu\nu})=0.\label{con}
\end{equation}
Here, $\hat\nabla_\beta$ denotes the covariant derivative calculated with respect to $\hat\Gamma$. Notably, by defining a new metric tensor as:
\begin{equation}\label{met}
\hat{g}_{\mu\nu}=f'(\mathcal{R})g_{\mu\nu},
\end{equation}
we can rewrite Eq. \eqref{con} as:
\begin{equation}\label{met2}
\nabla_\beta(\sqrt{- \hat{g}} \hat{g}^{\mu\nu})=0,
\end{equation}
provided that the independent connection $\hat\Gamma$ is the Levi-Civita connection associated with the metric $\hat g_{\mu\nu}$. It is also evident that for linear $f$, the connection $\hat\Gamma$ reduces to the Levi-Civita connection $\Gamma$ associated with the metric $g_{\mu\nu}$.

We can gain further insight into the properties of this theory of gravity by analyzing the $g$-trace of Eq. \eqref{structural}:
\begin{equation}
f'(\mathcal{R})\mathcal{R}-2 f(\mathcal{R})=\kappa^2 T,\label{struc}
\end{equation}
where $T$ represents the trace of the energy-momentum tensor. This expression reveals that $\mathcal{R}$ can be algebraically expressed as a function of matter fields for a given $f(\mathcal{R})$. In the case of vacuum and/or pure radiation spacetime, where $T=0$, we can solve Eq. \eqref{struc}, which implies that the theory reduces to GR with a cosmological constant. Furthermore, it can also be demonstrated that the theory does not possess any additional degrees of freedom beyond those present in GR. This is in contrast to the case of metric $f(R)$ gravity, where the relationship between $R$ and $T$ arises from a differential equation, introducing an extra degree of freedom (which can be interpreted as a scalar field with appropriate transformations).

In the following analysis, we will consider the simplest extension of GR, often referred to as the Starobinsky or quadratic model, given by:
\begin{equation} \label{Eq:fR Quadratic}
f(\mathcal R) = \mathcal R + \beta \mathcal R^2,
\end{equation}
where $\beta$ is a parameter. In this case, the solution of Eq. \eqref{struc} takes the form of GR, i.e., $\mathcal R = -\kappa^2 T$. This implies that any modifications appearing in the field equations are functions of the matter fields, parametrized by a single parameter $\beta$.

Once we have chosen a specific model of gravity, we can derive the Tolman-Oppenheimer-Volkoff (TOV) equation, which describes the equilibrium of a compact stellar object modeled as a perfect fluid in the framework of Palatini gravity with an arbitrary $f(\hat R)$ function \cite{Wojnar:2017tmy}. However, for practical numerical calculations, the TOV equation can be rewritten in a more suitable form \cite{Kozak:2021vbm}. In this form, denoting the derivative with respect to the radial coordinate $r$ as $'$, the TOV equation reads:
\begin{align}\label{tovJ1}
        p' =  \Bigg[&-\frac{G\mathcal{M}(r)}{r^2\mathcal{I}^{1/2}_1}(\varepsilon+p)\left(1 - \frac{2G\mathcal{M}(r)}{r\mathcal{I}^{1/2}_1}\right)^{-1}
   \left(1 + \frac{4\pi\mathcal{I}_1^\frac{3}{2}r^3}{\mathcal{M}(r)}\left(\frac{p}{\mathcal{I}^2_1} + \frac{\mathcal{I}_2}{2\kappa^2}\right)\right)\Bigg]  \\
        &\times \left(\frac{r}{2}\partial_{r} \ln\mathcal{I}_1 + 1 \right)
   +\left(-\varepsilon + 5p\right)\partial_{r} \ln\mathcal{I}_1,\nonumber
\end{align}
 where the functions $\mathcal{I}_1$ and $\mathcal{I}_2$ in the case of a perfect-fluid stress-energy tensor are  ($c^2\rho=\varepsilon$):
\begin{align}
    & \mathcal{I}_1 =  1 + 4\beta \kappa^2 (\varepsilon - 3p) \label{Iota1}, \\
    & \mathcal{I}_2 = \frac{4\beta\kappa^4(\varepsilon - 3p)^2}{\left(1 + 4\beta \kappa^2 (\varepsilon - 3p) \right)^2} \label{Iota2}.
\end{align}
The effective mass function has the following form:
\begin{equation}\label{masaRel} 
\begin{split}
    \mathcal{M}(r) =    \int^{r}_0 4\pi\tilde{r}^2 \frac{\varepsilon - 2\beta\kappa^2(\varepsilon - 3p)^2}{\left(1 + 4\beta \kappa^2 (\varepsilon - 3p) \right)^{1/2}}
    \times\left[1+ \frac{\tilde{r}}{2}\partial_{\tilde{r}}\ln\left(1+4\beta\kappa^2 (\varepsilon -3p)\right)\right]d\tilde{r}.
\end{split}
\end{equation}

It is important to note that equations \eqref{tovJ1}-\eqref{masaRel} are exact and do not involve any approximation. They provide a complete description of the equilibrium of a compact stellar object in Palatini gravity with the specific $f(\mathcal R)$ function given by \eqref{Eq:fR Quadratic}. These equations reduce to the corresponding equations in GR when $\beta=0$. By setting $\beta=0$ in our analysis, we can compare and examine the differences between these two models of gravity using our EoSs.

{In the framework of $f(\hat R)=\hat R+\beta \hat R^2$ model,
 the weak-field limit analysis has shown that $|\beta|$ typically falls below $2\times 10^{8},\text{m}^2$ \cite{Olmo:2005zr}. However, due to uncertainties in microphysics, experiments within the Solar System have not been able to impose precise constraints on these parameters \cite{Toniato:2019rrd}. Note that tests of gravity considered in vacuum, as for instant Shapiro delay technique, do not provide constraints for Palatini gravity because the theory reduces to GR with the cosmological constant, as discussed after the equation \eqref{struc}.
On the other hand, when considering microphysical aspects, seismic data from Earth has imposed stricter limitations on the parameter, with $|\beta|\lesssim 10^9 ,\text{m}^2$ (to a $2\sigma$ level of accuracy) \cite{Kozak:2023axy,Kozak:2023ruu}. Notably, in non-relativistic limit studies, it has been established that only the quadratic term is significant, while higher-order terms, starting at the sixth order, become relevant \cite{Toniato:2019rrd}.
In a manner akin to general relativity, none of the $f(\hat R)$ models can effectively explain the rotation curves of galaxies \cite{Hernandez-Arboleda:2022rim,Hernandez-Arboleda:2023abv}. As a result, constraints from galaxy catalogs remain elusive to date.}

For convenience, we will redefine the Starobinsky parameter as:
\begin{equation}
\alpha := 2\kappa^2 \beta.
\end{equation}
This redefinition of the parameter ensures that $\alpha$ and $\beta$ have the same units (km$^2$), assuming $c=G=1$, since the Ricci tensor, energy density $\varepsilon$, and pressure $p$ are given in units of km$^{-2}$.

Before delving into a more detailed analysis of the gravity models, let us briefly discuss the singular values of the parameter $\alpha$. It is important to note that the TOV equation \eqref{tovJ1} and the mass equation \eqref{masaRel} become singular when $\mathcal{I}_1=0$. This singularity arises from the fact that these equations were derived using a conformal transformation, which only maps a subset of possible solutions \cite{Wojnar:2017tmy, Kozak:2021vbm}. Therefore, there exists a singular value of the parameter $\alpha$, denoted as $\alpha_s$, that depends on the trace of the stress-energy tensor:
\begin{equation}\label{singu}
\alpha_s=-\frac{1}{2(\varepsilon-3P)}.
\end{equation}
The singular value $\alpha_s$ is responsible not only for the unbounded behavior of the stellar mass \eqref{masaRel}, but also for the singular behavior of the TOV equation \eqref{tovJ1}.

Hence, in order to solve the equations \eqref{tovJ1} and \eqref{masaRel}, it is crucial to determine the allowed range of values for the parameter $\alpha$, excluding the singular value $\alpha_s$. Moreover, we should note that the singular value depends on the chosen EoS, meaning that each energy density can be associated with a specific $\alpha_s$. Therefore, it is necessary to establish a "healthy" range for the parameter $\alpha$ that corresponds to the expected range of central energy densities in the core of a neutron star (250 $\leq \varepsilon_c \leq 1500$ MeV/fm$^{-3}$). To accomplish this, we first need to determine the values of $\alpha_s$ for each EoS considered in this work.


\section{Results}\label{MGresults}

In the following analysis, we will consider three cases: interpolation starting at (1) the saturation density, (2) twice the saturation density, and (3) phase transitions at one and a half times the saturation density.

\subsubsection{Trace anomaly}

\begin{figure}
    \centering
    \includegraphics[width=0.495\textwidth]{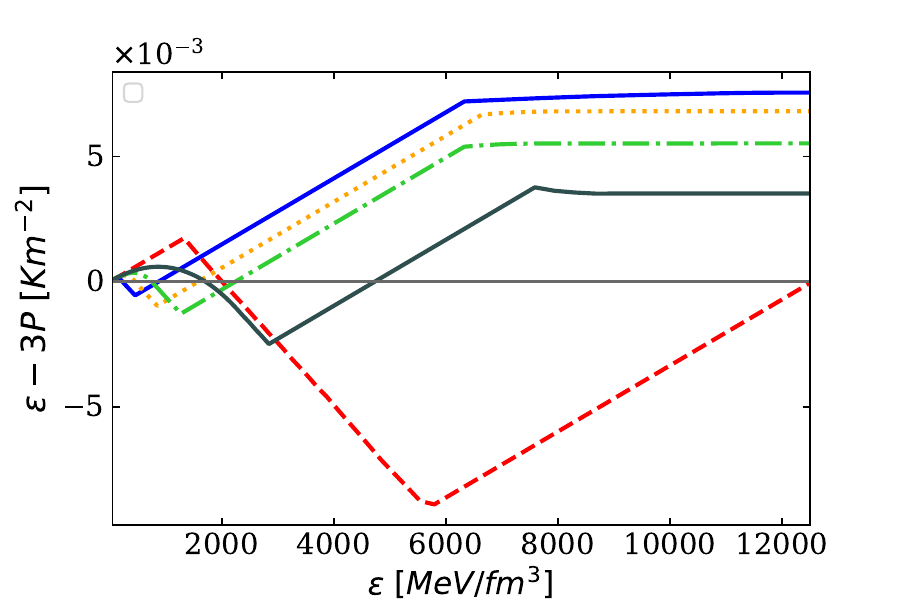} \includegraphics[width=0.495\textwidth]{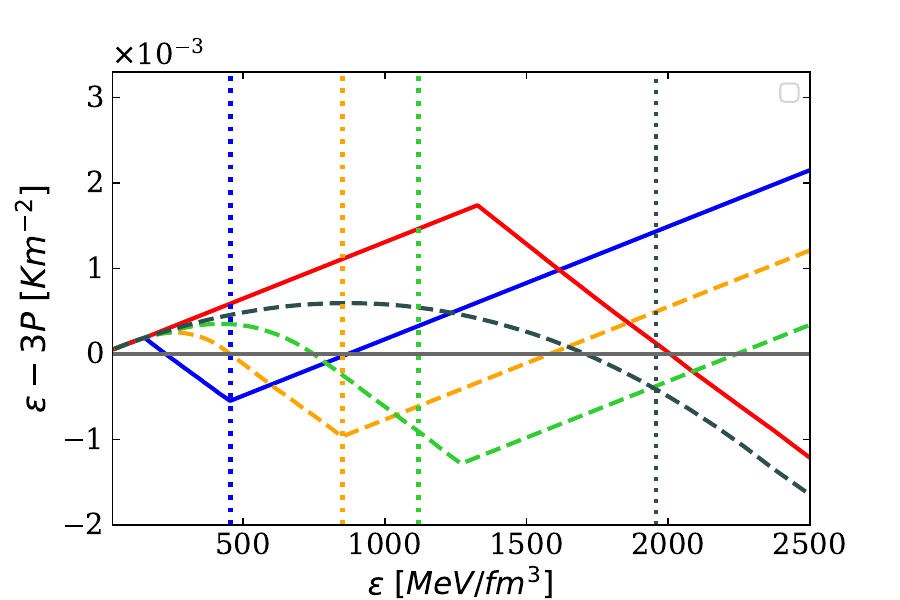}   
    \caption{\label{fig:TracevE}{\bf Left:} Trace ($\varepsilon-$3P) of the momentum-energy tensor as a function of  energy density for the five EoSs at $n=n_s$ of the Fig. \ref{fig:EoSinterpol2} for all the energy-density ranges. {\bf Right:} Trace values for the same EoSs at typical energy density ranges inside a NS (in GR). The vertical dotted lines stand for the corresponding central density energy for each one.}
\end{figure}

By analyzing the values of $\varepsilon-3P$ in Fig. \ref{fig:TracevE}, we observe that all the considered EoSs exhibit an energy density interval where the trace becomes negative, consistent with the results of \cite{Annala:2023cwx,Brandes:2023hma}. This behavior is dependent on the stiffness of each EoS. It is expected that this occurs in cases with high slopes, which are necessary for neutron stars to attain the observed high masses and radii in the framework of GR.

It is worth noting that although for $\mu_B \rightarrow \infty$ the trace $\varepsilon-3P \rightarrow 0^+$ due to pQCD reasoning for a relativistic asymptotically free gas, in a hadronic phase or a strongly repulsive quark phase (far from asymptotic freedom), the interaction can be very repulsive, leading to $P \geq \frac{\rho c_s^2}{3}$ \cite{Bedaque:2014sqa,Ecker:2022dlg,Roy:2022nwy,Fujimoto:2022ohj,Marczenko:2022jhl}.

Since this behavior is consistent with causality, it cannot be ruled out based on first principles. In practice, even in the framework of GR, it is quite possible for $\mathrm{Tr}\,T_{\mu\nu}< 0$ in certain intervals due to the stiffness of the EoS required to accommodate the maximum masses of neutron stars.

As a result, $\alpha_s$ as a function of the trace exhibits discontinuities at points where the trace changes sign, resulting in both positive and negative values. Due to this characteristic, it is necessary to explore a wide range of $\alpha$ values to avoid this feature within a NS.

Once we have determined a range of permissible values for $\alpha$, the next step is to select among them those values that yield M-R diagrams that satisfy the requirements imposed by astrophysical observables, specifically the observed masses and radii of neutron stars. There are usually extracted in the context of GR, but because this happens in a regime of small or null density outside the star, it can be accepted as a first approximation. 

Regarding the mass constraints, we consider the lower and upper bounds of the mass estimate for PSR J0952-0607 ($2.35 \pm 0.17$ M$\odot$) \cite{Romani:2022jhd}. For the radius constraints, we take into account the measurements obtained by the NICER experiment for neutron stars J0740+6620 \cite{Miller:2021qha, Riley:2021pdl} and J0030+0451 \cite{Miller:2019cac,Riley:2019yda}. As a result, we must discard EoSs that predict radii $R < 10.75$ km for $M= 2.0 M\odot$ and $R < 10.8$ km for $M= 1.1 M_\odot$ (at 1 $\sigma$ confidence) within the considered models of gravity. Additionally, it is worth noting that there are EoSs that do not satisfy the observational constraints in the context of GR, but may still satisfy them for specific values of $\alpha$ within the allowed range in Palatini gravity. 

\subsection{Interpolation at saturation density}
As it turns out, matching with the chiral band at the saturation density $n=n_s$ provides the widest range of interpolation for the EoS. Consequently, this band plays a crucial role in determining the upper bounds for the Starobinsky parameter $\alpha$ in NS. By considering the constraints imposed by the chiral band only up to $n_s$, we can establish the maximum values of $\alpha$ that are compatible with the observed properties of neutron stars.

\subsubsection{The singular Starobinsky parameter}
\begin{figure}
    \centering

    \includegraphics[width=0.48\textwidth]{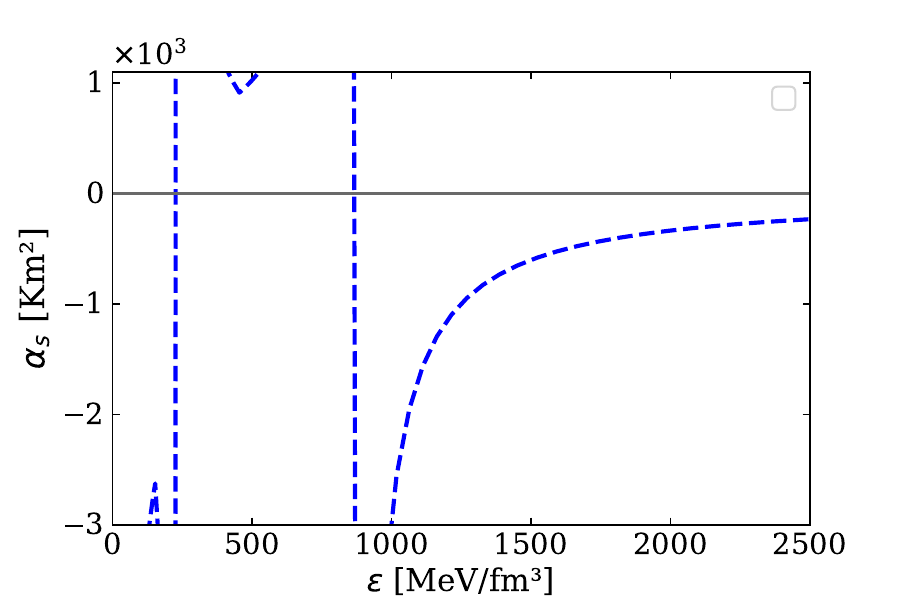}    \includegraphics[width=0.48\textwidth]{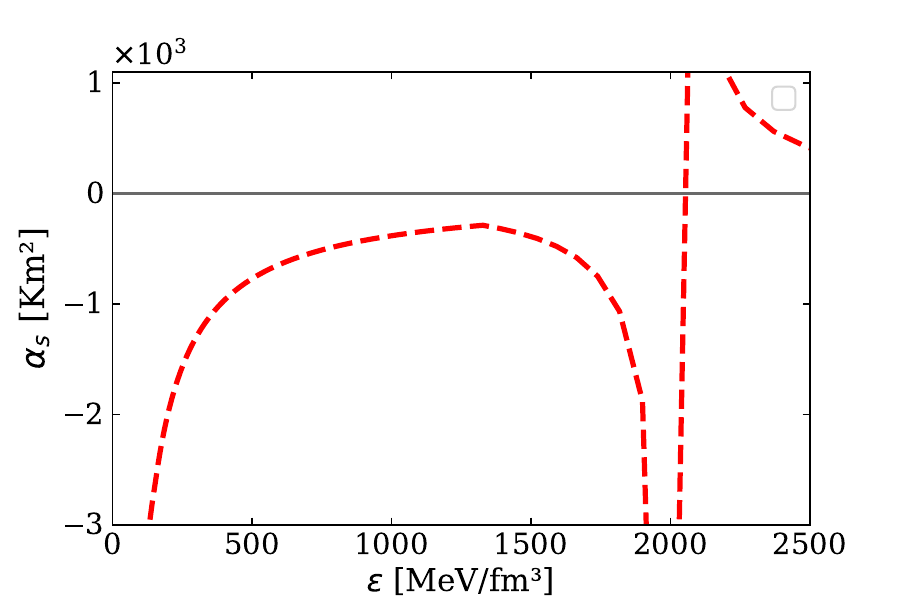}
    \caption{ \label{fig:maxminrange} Values of the singular parameter $\alpha_s$ for which $M \rightarrow \infty$, for the stiffest (Blue EoS) and softest (Red EoS) EoS obtained at matching point $n=n_s$ with $\chi$EFT EoS (top of Fig. \ref{fig:EoSinterpol2}). {These later figures show that for the stiffest EoS, the allowed values of  $\alpha_s$ in the energy density range $50 < \varepsilon < 2500$ MeV/fm$^3$  is $-230 \lesssim \alpha \lesssim 915$ km$^2$. However,  considering smaller values of the energy density  ($\varepsilon \lesssim 400$ MeV/fm$^3$), the allowed values of  $\alpha_s$ is $-2600 \lesssim \alpha \lesssim 1100$ km$^2$. For the softest EoS, $-300 <\alpha \lesssim 420$  km$^2$.} Note that taking into account  another ranges of energy-density, $\alpha_s$ can overcome these values.}
   \end{figure}

As mentioned, in order to apply modified gravity using the TOV equations \eqref{tovJ1} and \eqref{masaRel}, we need to identify the singular values of the parameter $\alpha_s$.

As a general criterion, we take the range of values of $\alpha$ between the maximum of the negative $\alpha_s$ and the minimum of the positive $\alpha_s$. {Thus, from the Fig. \ref{fig:maxminrange}, for energy densities in the range $50 < \varepsilon < 2500$ MeV/fm$^3$, the parameter $\alpha$ can take the values $-230 \lesssim \alpha \lesssim 915$ km$^2$ for the stiffest EoS, and  $-300 < \alpha \lesssim 420$  km$^2$ for the softest one. However, for smaller energy densities, $\alpha_s$ can overcome these values. Thus, for  $\varepsilon  <$ 400 MeV/fm$^3$ we deal with $  -2600<\alpha_s<1100$ while for the stiffest EoS we have $\alpha> -1000 $ for the softest EoS. Let us notice that we do not know what is a value of the central density of a neutron star, therefore we have decided to consider various possible ranges as dictated by the singular parameter $\alpha_s$. As a result, in order to account for different values of the central energy density in a NS, a starting point to look for values of the Starobinsky parameter lies in the approximate range $-2000 < \alpha \lesssim 1100$ km$^2$.}

\subsubsection{Constraining Starobinsky parameter}
Let us now discuss the results obtained for different values of the parameter $\alpha$, using our five EoSs shown in Fig. \ref{fig:EoSinterpol2}, along with the observational constraints on mass and radius. In all cases, the GR scenario corresponds to $\alpha=0$ and is represented by a black curve. The mass constraint is indicated by the grey shaded area, while the radius constraint ($R=10.75$ km) is shown as a dash-dotted gray line. Additionally, we have marked two pulsars measured by NICER: PSR $J0740+6620$ (represented by a red circle) and PSR $J0030+0451$ (represented by a cyan circle).

\begin{figure}
    \centering    \includegraphics[width=0.80\textwidth]{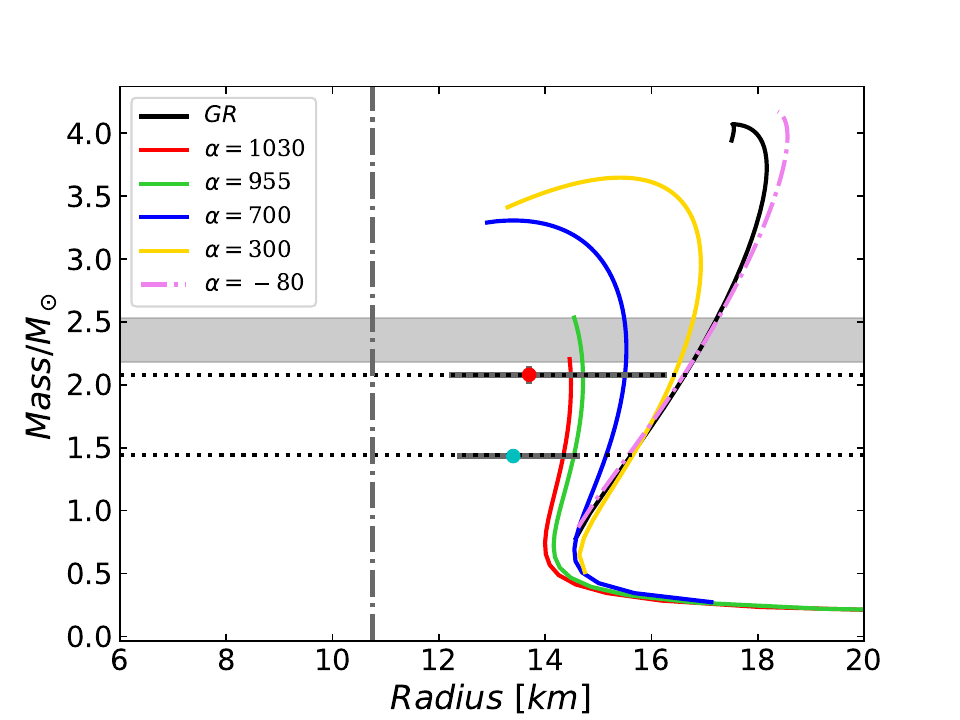}    \includegraphics[width=0.49\textwidth]{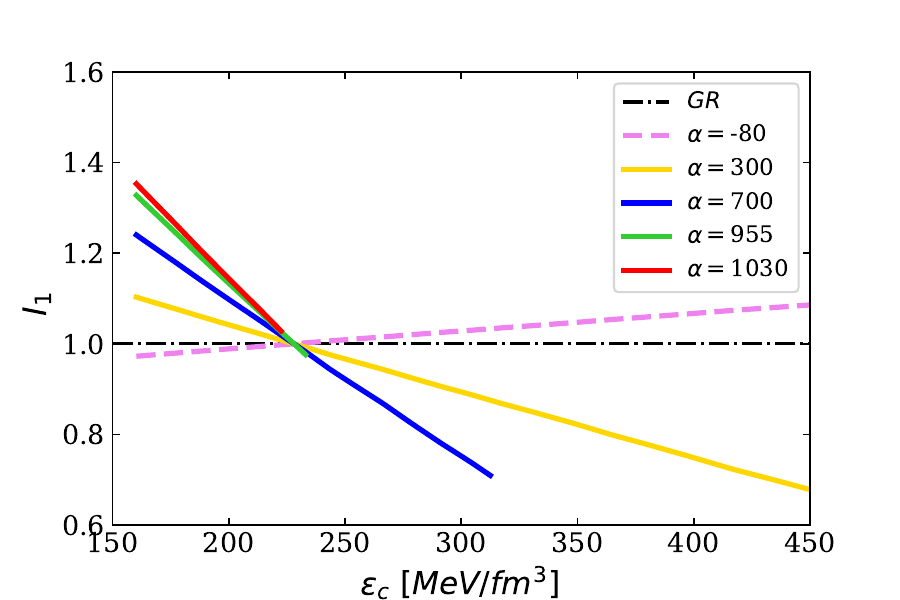}     \includegraphics[width=0.49\textwidth]{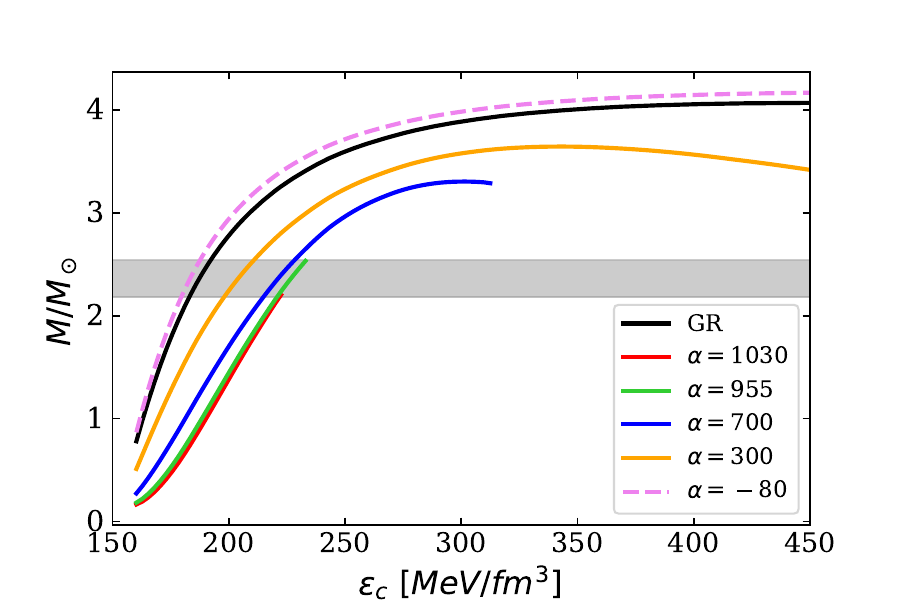}\\     \includegraphics[width=0.49\textwidth]    {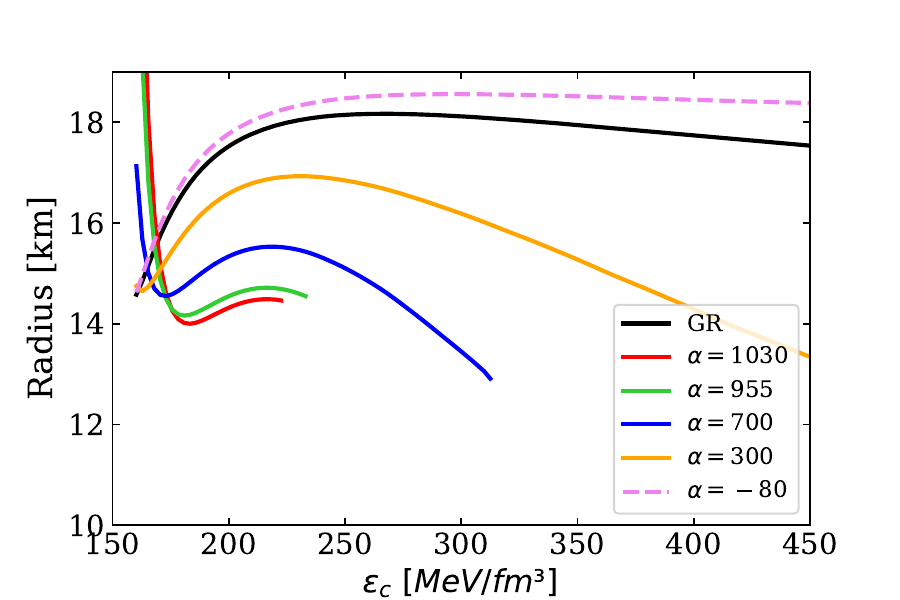}     \includegraphics[width=0.49\textwidth]{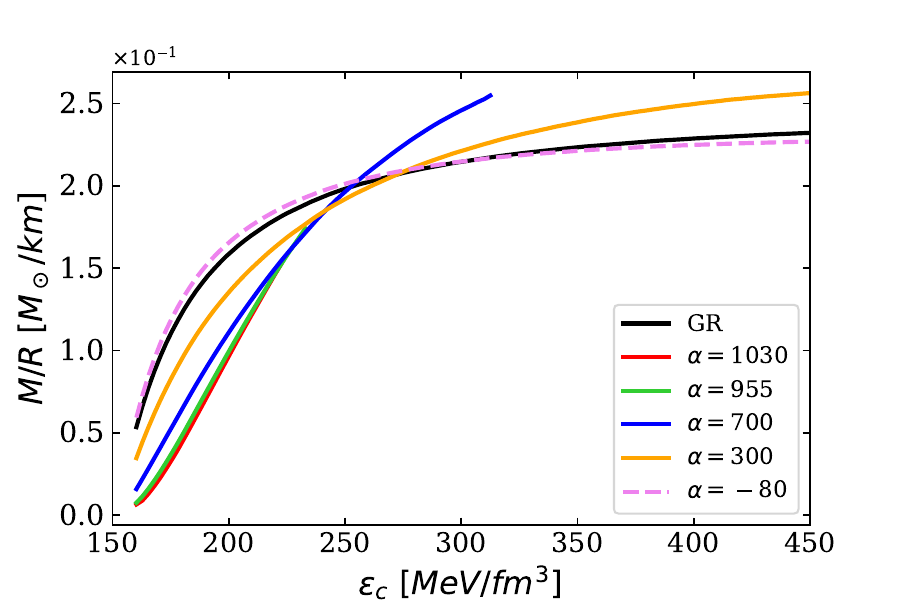}
    \caption{{\bf Top:} Mass-Radius diagrams in GR (black solid line) and ETG (coloured lines) for the maximally stiff (Blue) EoS of Fig. \ref{fig:EoSinterpol2}). The grayish shaded area contains the  lower and  upper mass bound (2.17 and 2.52 $M_\odot$, respectively) and the vertical dash-dotted line indicates the    lower radius bound (10.8 km). {\bf Second row:} $\mathcal{I}_1$  (left) and mass (right) against the central energy density for Blue EoS for different values of $\alpha$. {\bf Third row:} Radius (left) and compactness (right) against the central energy density for Blue EoS for different values of $\alpha$. }    
    \label{fig:MvREoSmax}
\end{figure}

Considering the Blue EoS (the stiffest) presented in Fig. \ref{fig:MvREoSmax}, we find that the lowest maximum mass constraint reached by the mass-radius (M-R) curves is obtained for $\alpha=1030$ km$^2$ (represented by the red color). In the case of the Blue EoS, the maximum mass and radius obtained in the GR framework are $4.07 M_\odot$ and $17.5$ km, respectively. It is important to note that we only consider positive values of $\alpha$ since the maximum mass does not fall within the observational limits for negative values. The positive values of $\alpha$ lead to a reduction in both mass and radius compared to the results obtained in the GR framework. Hence, it is not surprising that very high values of $\alpha$ are required to significantly decrease these bulk parameters.

The range of values that satisfy both the mass and radius constraints\footnote{In the case of the radius, the curves lie in the $1\sigma$ confidence level according to the NICER mission. The radius constraint is not strict, however, in this very first approach we use it to bound the parameter.} for the Blue EoS is given by $955 \leq \alpha \leq 1030$. This means that $\alpha=1030$ corresponds to the lower mass bound (represented by the red line), while $\alpha=955$ corresponds to the upper mass bound (represented by the green line). Since the Blue EoS is the stiffest among the considered EoSs, this range can be considered as an upper bound for the positive values of the Starobinsky parameter in NS.

\begin{figure}
     \centering
     \includegraphics[width=0.80\textwidth]{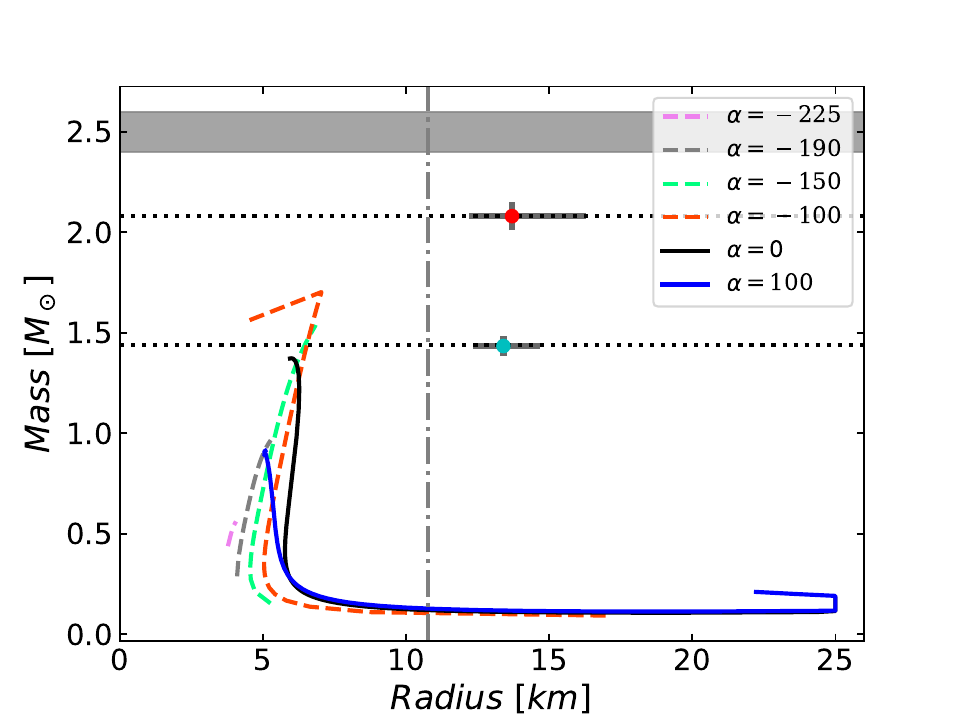}
      \caption{Mass-Radius diagrams in GR (black solid line) and ETG (coloured lines) for the softest (Red) EoS of \ref{fig:EoSinterpol2}. The grayish shaded area refers to the lower and  upper mass bound (2.17 and 2.52 $M_\odot$, respectively) and the vertical dash-dotted line indicates the lower radius bound ($\approx$ 10.8 km). These very soft EoS are ruled out by the data in both GR and Palatini f($\mathcal R)$}
      \label{fig:MvREoSmin}
 \end{figure}

The Red EoS (the softest) in Fig. \ref{fig:MvREoSmin} provides the opposite case to the Blue EoS. Based on the values of $\alpha_s$ shown in the bottom-right panel of Fig. \ref{fig:maxminrange}, the range of allowed values for this parameter must be negative. However, the M-R sequences for the Red EoS are less affected by the considered ETG compared to GR, and they do not reach the required mass and radius bounds.

This is so because for this particular EoS, the solution of the TOV equation within the GR framework yields neutron star masses for a central energy density range $1330 < \varepsilon_c < 4100$ MeV/fm$^3$. In this range, the allowed values of $\alpha$ are small ($\alpha>$ -250 km$^2$), and therefore these values are not sufficient to produce significant increases in mass and radii.

Although this EoS is not able to provide values of $\alpha$ that satisfy the observational constraints, it is still useful in establishing a negative range for $\alpha$. This negative range can then serve as a starting point for searching for parameter values in other soft EoSs.

\begin{figure}
      \centering      \includegraphics[width=0.80\textwidth]{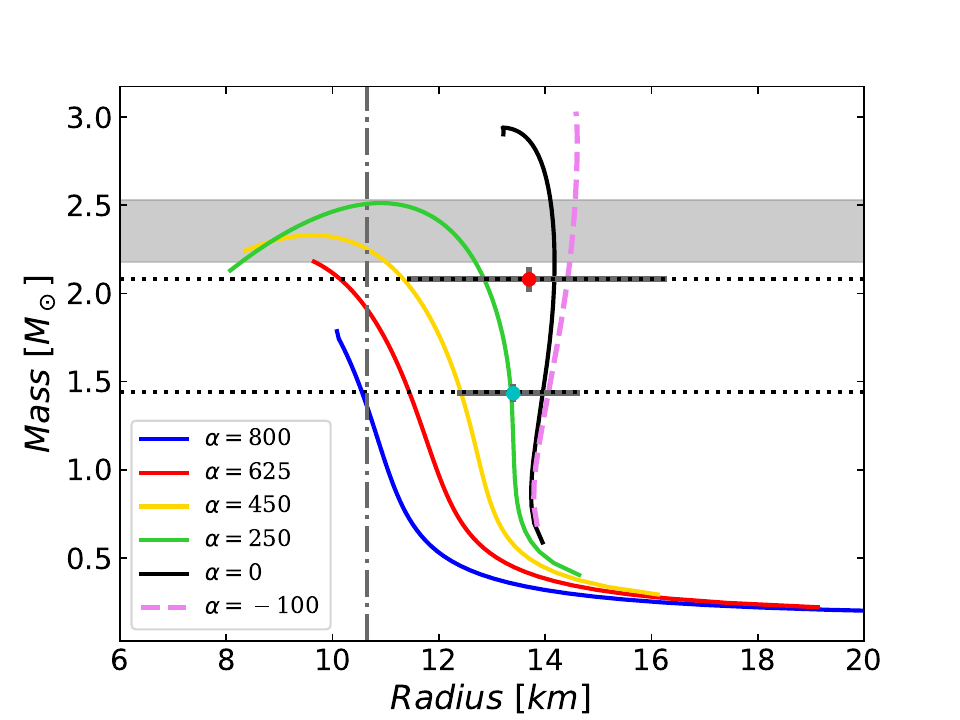}
      \includegraphics[width=0.45\textwidth]{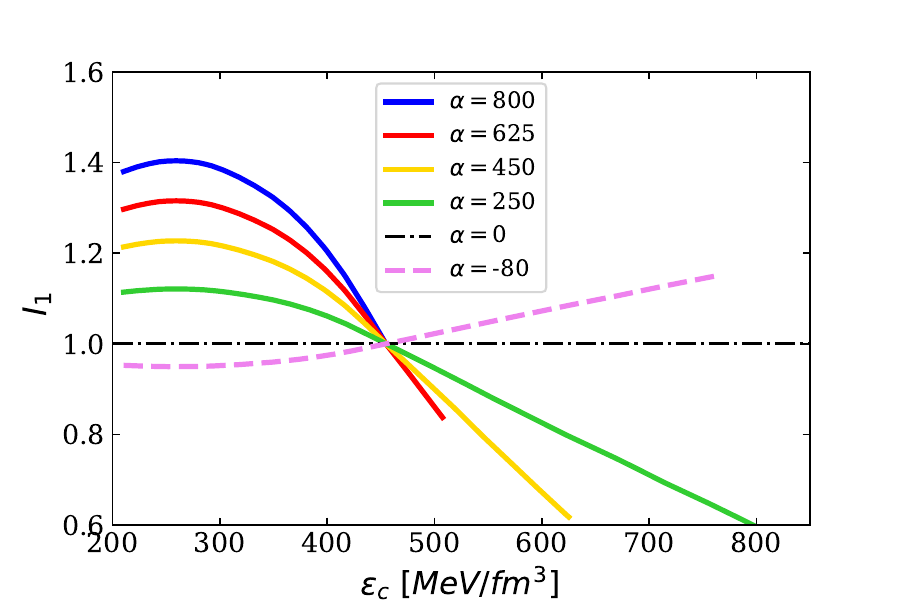}
     \includegraphics[width=0.45\textwidth]{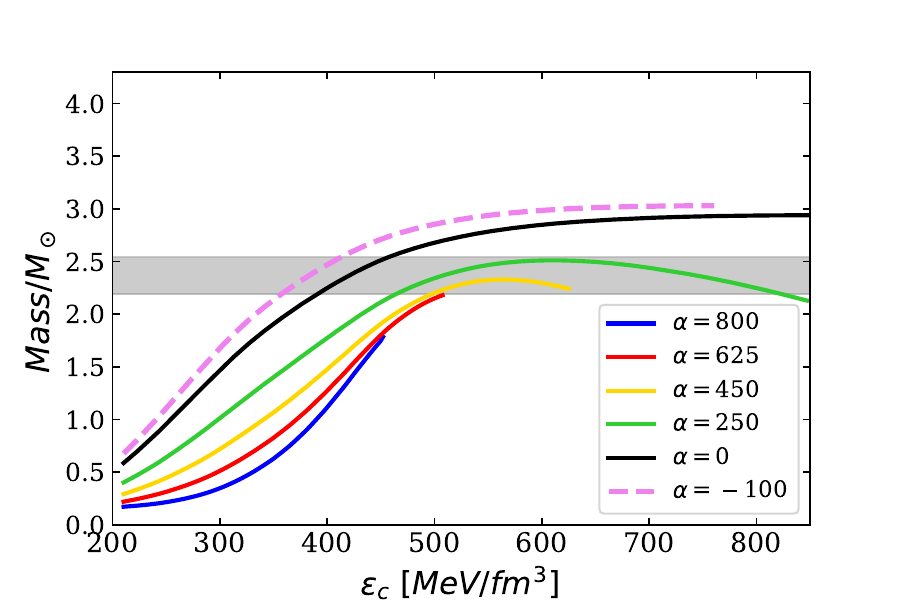}\\
     \includegraphics[width=0.45\textwidth]{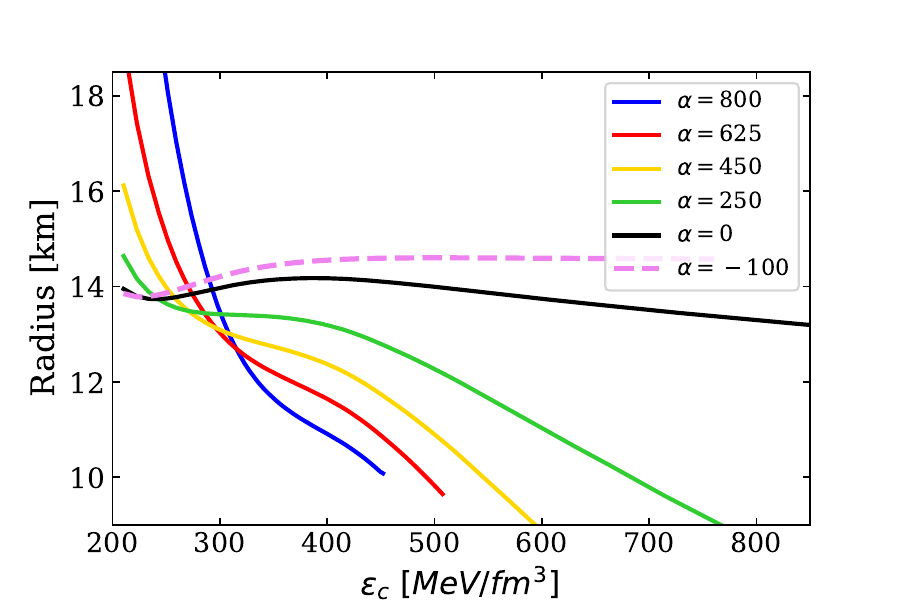}
     \includegraphics[width=0.45\textwidth]{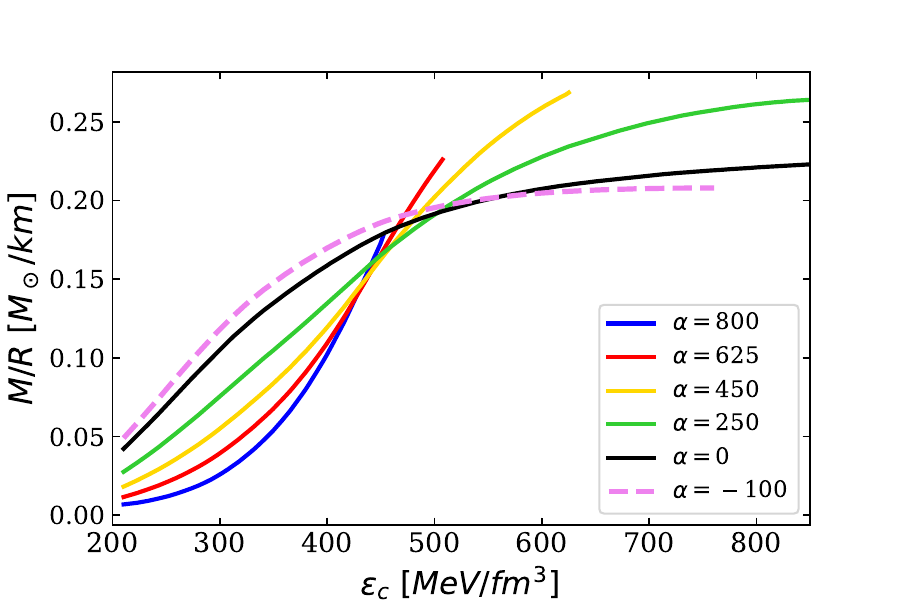}
      \caption{{\bf Top:} Mass-Radius diagrams in GR (black solid line) and ETG (coloured lines) for the second stiffest (Orande) EoS of \ref{fig:EoSinterpol2}. The grayish shaded area refers to the lower and  upper mass bound (2.17 and 2.52 $M_\odot$, respectively) and the vertical dash-dotted line indicates the lower radius bound (10.8 km). {\bf Second row:}  Mass (left) and radius (right) against the central energy density  for different values of $\alpha$. {\bf Third row:}  Mass and compactness against  the central energy density  for different values of $\alpha$.}
      \label{fig:MvREoSinterm1}
  \end{figure}

Let's analyze the Orange EoS of Fig. \ref{fig:EoSinterpol2} (up), which is quite stiff, based on the M-R results in Fig. \ref{fig:MvREoSinterm1}. The range of allowed values for $\alpha$ is between positive values that lead to decreasing mass and radius.

This EoS has a maximum mass of 2.94 M$\odot$ within the GR framework ($\alpha = 0$), which is larger than the upper mass bound of 2.52 M$\odot$. Therefore, we need positive values of $\alpha$ to decrease the maximum mass and radius to values below this upper bound.

From the M-R curves, we find that the lower mass bound within the allowed radius is reached for $\alpha = 625$ km$^2$ (solid red line), while the upper bound is reached for $\alpha = 250$ km$^2$. Assuming that this EoS represents the true equation of state of a neutron star, the allowed values for the parameter $\alpha$ would lie within the range $250 \leq \alpha \leq 625$ km$^2$.

\begin{figure}
     \centering
     \includegraphics[width=0.80\textwidth]{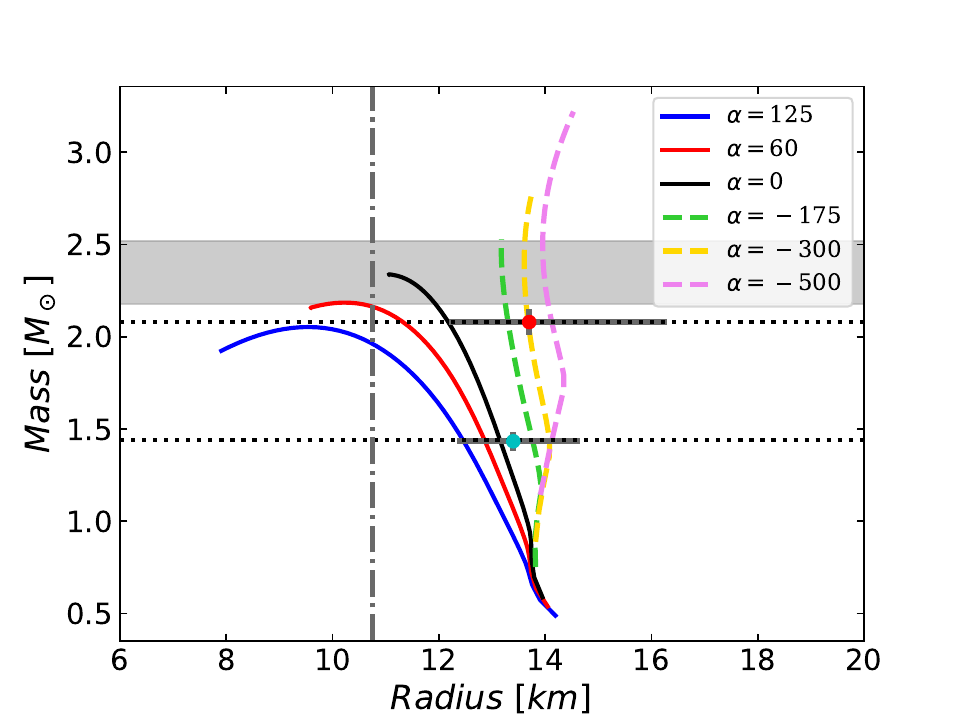}         \includegraphics[width=0.45\textwidth]{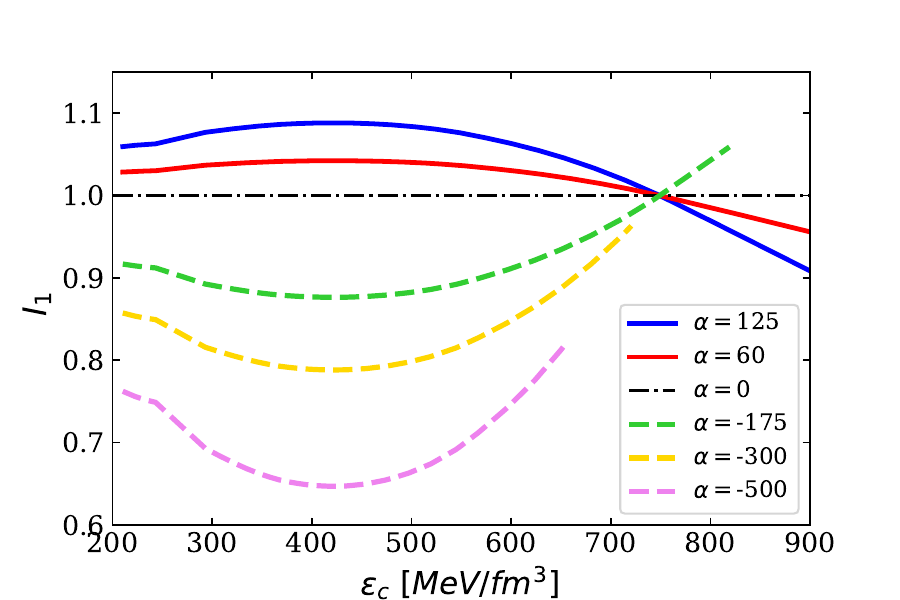}
     \includegraphics[width=0.45\textwidth]{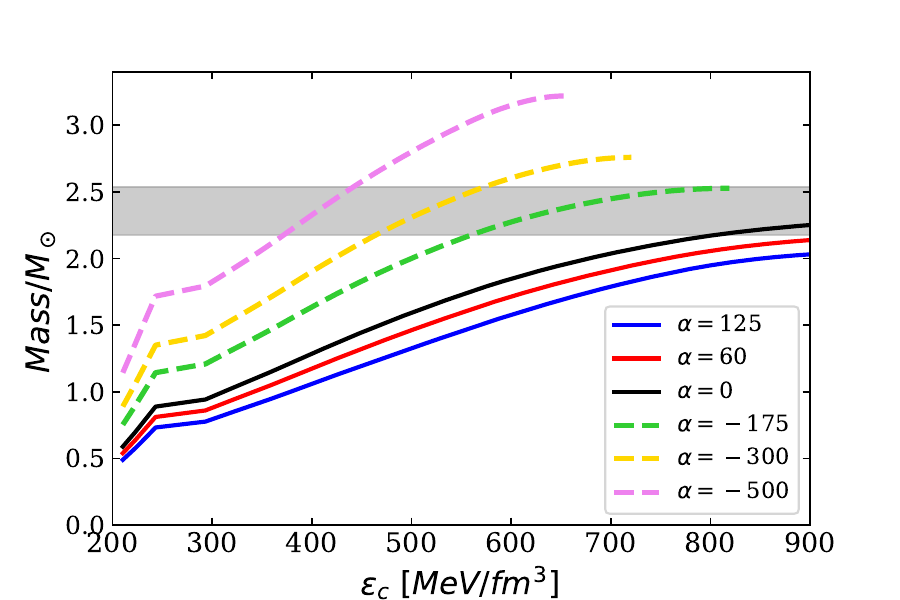}\\
     \includegraphics[width=0.45\textwidth]{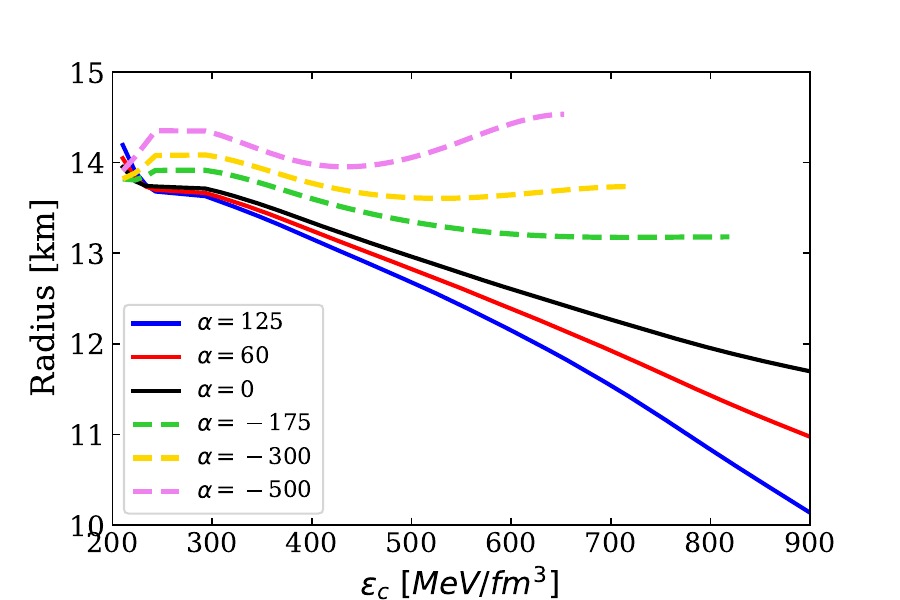}
     \includegraphics[width=0.45\textwidth]{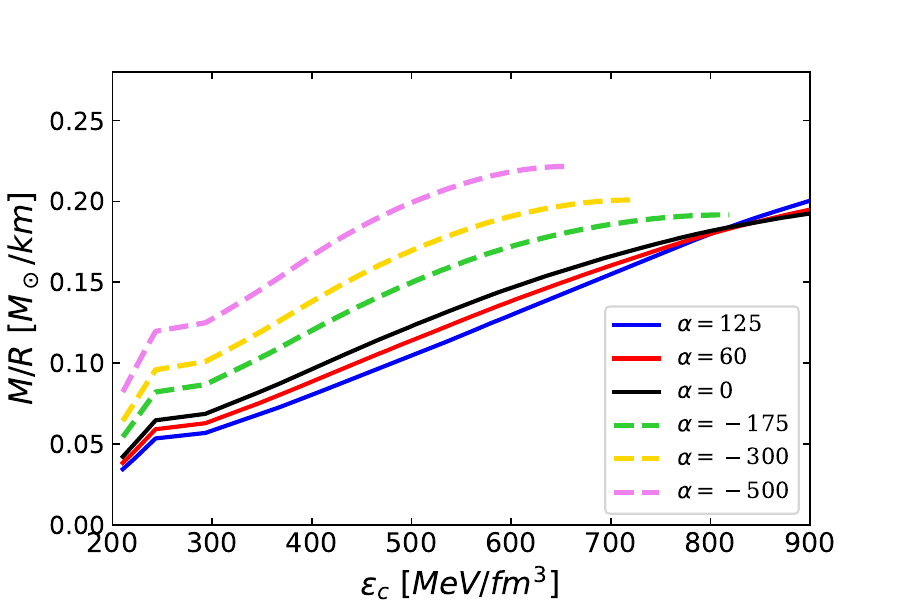}
      \caption{{\bf Top:} Mass-Radius diagrams in GR (black solid line) and ETG (coloured lines) (coloured lines) for the Green EoS of Fig. \ref{fig:EoSinterpol2}. The grayish shaded area represents the lower and upper mass bound (2.17 and 2.52 $M_\odot$, respectively) and the vertical dash-dotted line indicates the lower radius bound ($\approx$ 10.8 km). {\bf Second row:} $\mathcal{I}_1$ (left) and mass (right) aginst the  central energy density  for different values of $\alpha$. {\bf Third row:} Radius and compactness versus the central energy density  for different values of $\alpha$. }
      \label{fig:MvREoSintermPT}
 \end{figure}

Let's analyze the Green EoS of Fig. \ref{fig:EoSinterpol2} (up), which is also stiff, based on the M-R results in Fig. \ref{fig:MvREoSintermPT}. In the GR case, the maximum mass is reached at the midpoint of the shaded area between the mass limits (specifically, at $M=2.34$ M$_\odot$ and $R=11.06$ km for $\varepsilon_c = 1120$ MeV/fm$^3$).

To reach the upper mass bound of 2.52 M$\odot$, we need to consider negative values of the Palatini gravity parameter ($\alpha = -175$ km$^2$) in order to increase the star's radius and mass (indicated by the dashed green line in Fig. \ref{fig:MvREoSintermPT}). On the other hand, to reach the lower mass bound of 2.18 M$\odot$, positive values of $\alpha$ must be considered, specifically $\alpha = 60$ km$^2$ (indicated by the solid red line). This leads to a decrease in both mass and radius.

Therefore, the resulting range of allowed values for $\alpha$ dictated by this EoS is $-175 \leq \alpha \leq 60$ km$^2$.

\begin{figure}
    \centering
   \includegraphics[width=0.90\textwidth]{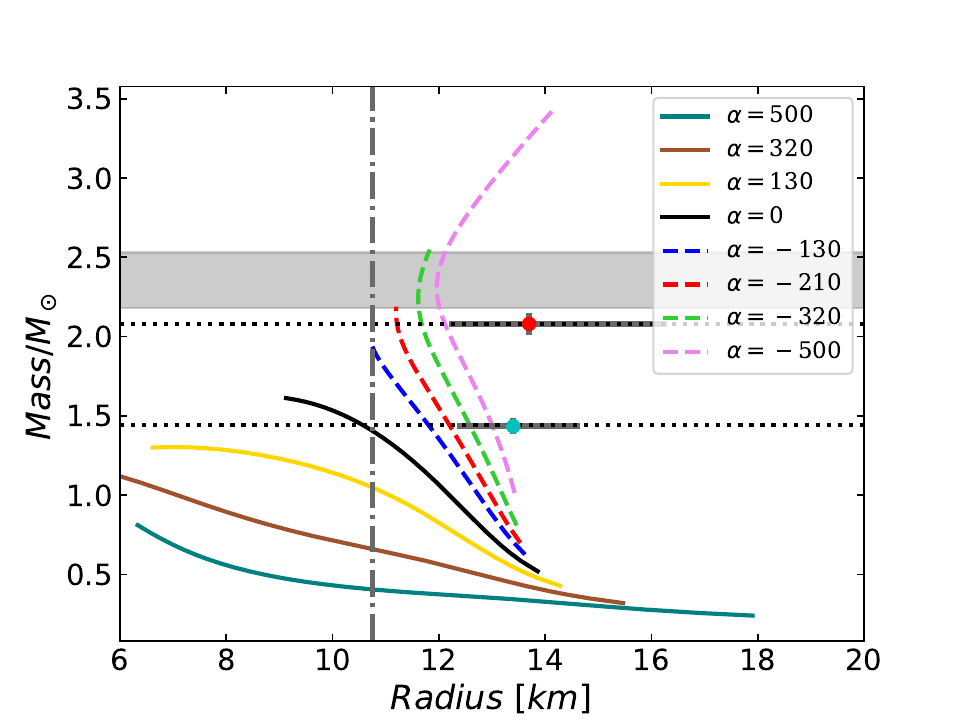}
     \includegraphics[width=0.48\textwidth]{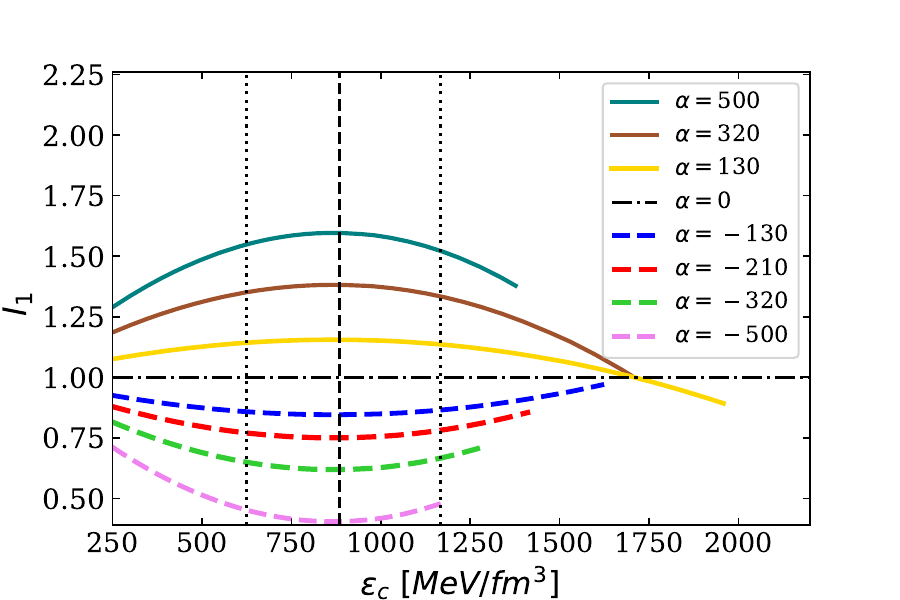}
     \includegraphics[width=0.48\textwidth]{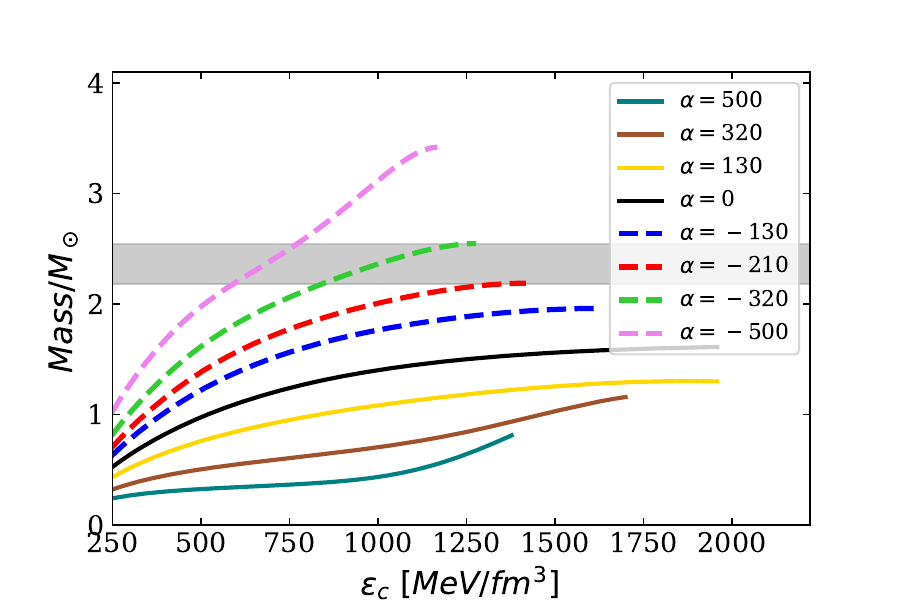}\\
     \includegraphics[width=0.48\textwidth]{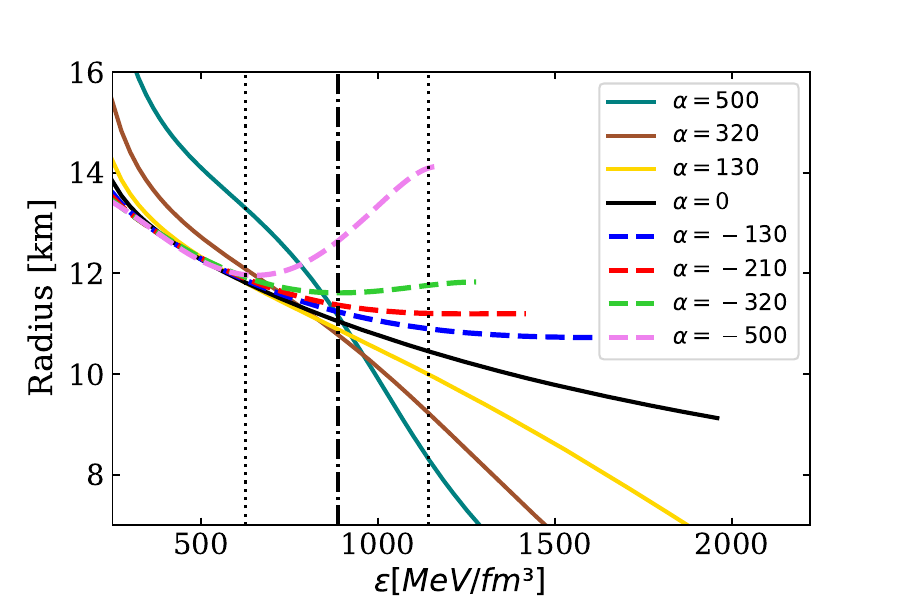}
     \includegraphics[width=0.48\textwidth]{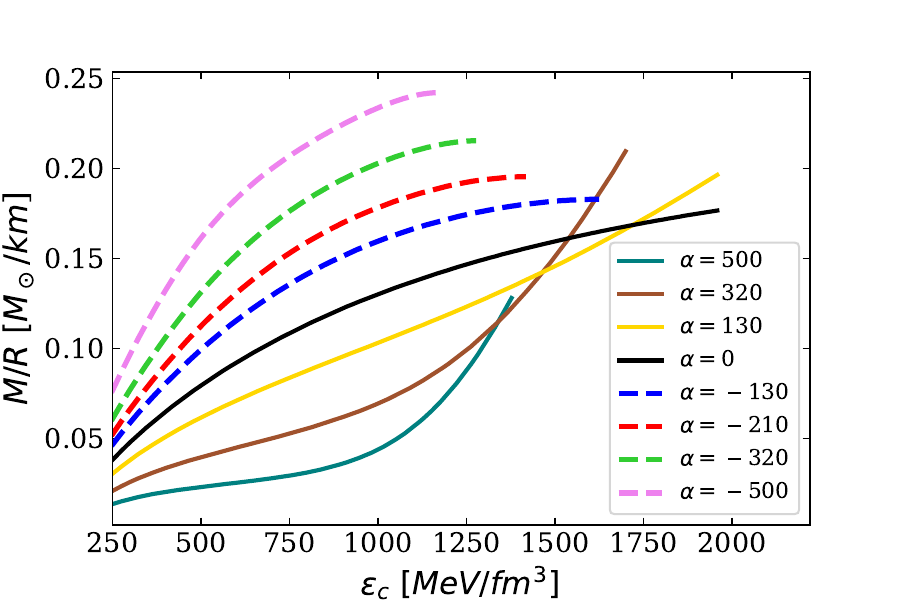}
      \caption{ {\bf Top:} Mass-Radius diagrams in GR (black solid line) and ETG (coloured lines) (coloured lines) for the soft (Grey) EoS of \ref{fig:EoSinterpol2}. The grayish shaded area refers to the  lower and  upper mass bound (2.17 and 2.52 M$_\odot$, respectively) and the vertical dash-dotted line indicates lower radius bound (10.8 km). {\bf Second row:} $\mathcal{I}_1$ (left) and mass (right) against the  central pressure  for different values of $\alpha$. The dashed vertical line indicates the $\varepsilon_c$ point in which $\mathcal{I}_1$ reaches the maximum/minimum value for positive/negative $\alpha$.  {\bf Third row:} Radius and compactness against  the central energy density  for different values of $\alpha$.}
      \label{fig:MvREoSintermlow}
\end{figure}


In the case of the Grey EoS, which is soft, the maximum mass reached in GR is below the lower mass limit ($M=1.61$ M$_\odot$, $R=9.12$ km for $\varepsilon_c = 1958$ MeV/fm$^3$). Therefore, in order to increase the star's mass and radius, negative values of $\alpha$ are required. This is illustrated by the red and green dashed lines in Figure \ref{fig:MvREoSintermlow}.

By considering the astrophysical observations, the resulting range of allowed values for this EoS is $-320 \leq \alpha \leq -210$ km$^2$. These negative values of $\alpha$ allow the EoS to reach the mass and radius limits imposed by the observational constraints.

In addition, we have also presented the results for all of our EoSs, excluding the Red EoS as it does not yield models that satisfy the observational constraints. These results include the variation of the function $\mathcal{I}_1$, stellar radius, mass, and compactness as functions of central density. The corresponding figures can be found in Fig. \ref{fig:MvREoSmax}, \ref{fig:MvREoSinterm1}, \ref{fig:MvREoSintermPT}, and \ref{fig:MvREoSintermlow}.

The function $\mathcal{I}_1$, depicted in the figures on the left side of the second row, exhibits a decrease/increase in its value with increasing central energy density for positive/negative values of $\alpha$, respectively. In all the curves, the value of $\mathcal{I}_1$ intersects with 1, which corresponds to General Relativity (GR), at the point where the trace of the stress-energy tensor changes sign. 

The plots of radius as a function of central energy density exhibit a strong correlation with the behavior of $\mathcal{I}_1$ for each EoS. However, it is important to note that this behavior differs for positive and negative values of $\alpha$.
For stiff EoSs, the relation between radius and energy density changes with the sign of $\alpha$. When positive values of $\alpha$ are allowed, the radius initially decreases as the energy density increases. However, at a certain point where the trace becomes negative, the radius starts to increase within a small range of energy density values before decreasing again. This behavior occurs at the point where the modified gravity effects become significant.
On the other hand, for less stiff EoSs with negative values of $\alpha$, a different pattern emerges. As the energy density increases, the radius also increases up to a particular value, beyond which the radius remains relatively constant even with further energy density increases. In this case, where $\alpha$ takes small negative values, although the trace changes sign, $\mathcal{I}_1$ remains practically constant with slight variations around 1.

In the case of the soft EoSs, the radius exhibits a decreasing trend with energy density for all values of $\alpha$. However, it is worth noting that the rate of decrease becomes more pronounced as positive values of $\alpha$ increase. Conversely, for negative values of $\alpha$, the radius decreases initially but reaches a point where it becomes approximately constant, independent of further increases in energy density\footnote{It is important to emphasize that these observations hold within the allowed ranges of $\alpha$ that satisfy the observational constraints.}. 

The plots of mass versus central energy density, depicted on the right side of the second row of Fig. \ref{fig:MvREoSmax}, \ref{fig:MvREoSinterm1}-\ref{fig:MvREoSintermlow} illustrate the increasing trend of mass with energy density. This growth in mass becomes more significant as the value of $\alpha$ decreases. The same pattern is observed in the plots of compactness versus central energy density, shown on the right side of the third row. The compactness is higher for lower values of $\alpha$, indicating a stronger gravitational field. However, this behavior changes for energy density values at which the trace becomes negative. In this region, the compactness becomes higher as the value of $\alpha$ increases.


\subsection{Interpolation at twice the saturation density}

\begin{figure}
    \centering
    \includegraphics[width=0.8\textwidth]{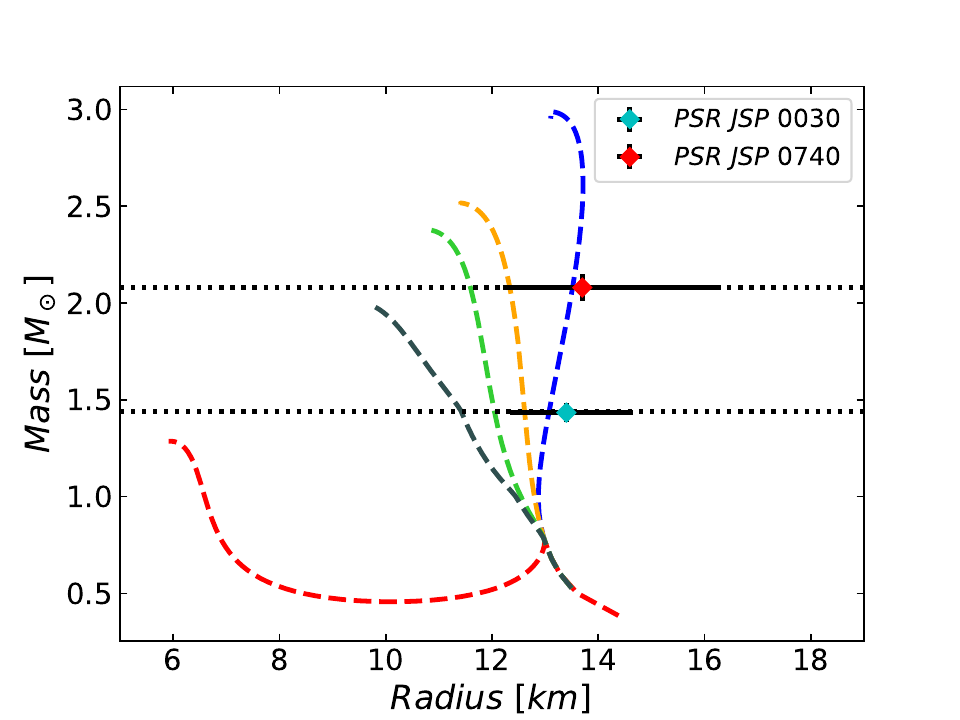}
    \caption{Mass-Radius diagrams for EoSs shown at the bottom of Fig. \ref{fig:EoSinterpol2} for GR, depicting the mass and radius values of two recently observed pulsars. The horizontal dotted lines represent masses of 1.4 $M_\odot$ and $2.1 M_\odot$, respectively}
    \label{fig:EoSMG}
\end{figure}

The first observation is that matching the chiral EoS at 2 times the saturation density significantly reduces the available parameter space compared to the interpolation at $n=n_s$. This is demonstrated in Figure \ref{fig:EoSinterpol2}, where the yellow and grey bands represent the interpolation at $n=n_s$, while the pale red and light blue bands represent the interpolation at $n=2n_s$. However, the interpolation at $n=2n_s$ ensures that the constructed EoSs satisfy the chiral requirements between $1$ and $2n_s$, which implies that they exhibit reasonable nuclear properties in this density range. Additionally, these EoSs usually satisfy GR astrophysical constraints for a canonical neutron star of $1.4 M_\odot$, particularly regarding tidal deformability and radius.

In the following analysis, we will use four EoSs from the previous section (Fig. \ref{fig:EoSinterpol2}), but we will replace the Grey EoS with a softer EoS that incorporates three intermediate first-order phase transitions (FoPT). We refer to this new EoS as the Grey1st EoS. The set of all five EoSs is shown in Fig. \ref{fig:EoSinterpol2} (bottom).

The Grey1st EoS features three FoPTs occurring at densities within the intervals of (1.92 - 2.20) $n_s$ (P=19.37 MeVfm$^{-3}$  and  $330.5 \leq \varepsilon \leq 348$ MeVfm$^{-3}$), (2.73-3.03) $n_s$ (P=38.8 MeVfm$^{-3}$ and $438 \leq \varepsilon \leq 491$ MeVfm$^{-3}$ )  and (3.98-4.53) $n_s$ (P=123.5 MeVfm$^{-3}$  and $666 \leq \varepsilon \leq 775$ MeVfm$^{-3}$), respectively. At the end of each FoPT, we use the same slope that the EoS had at the beginning of the transition.


\subsubsection{The singular Starobinsky parameter}

\begin{figure}
    \centering
    \includegraphics[width=0.45\textwidth]{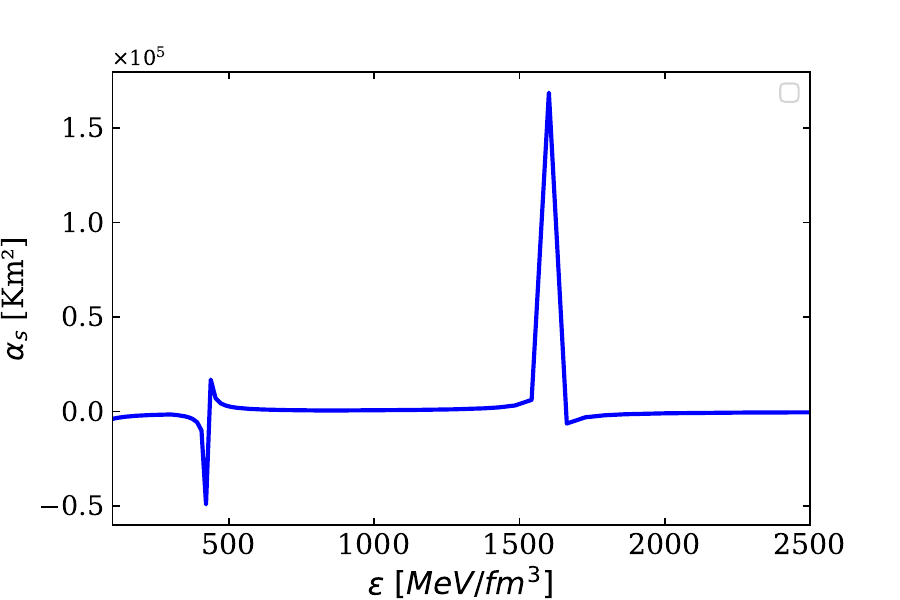}
    \includegraphics[width=0.45\textwidth]{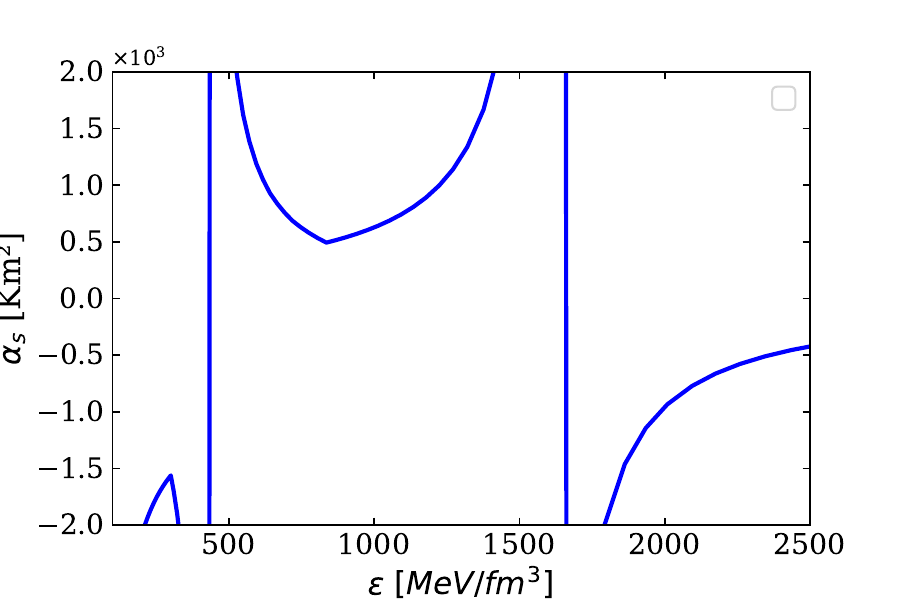}\\
        \includegraphics[width=0.48\textwidth]{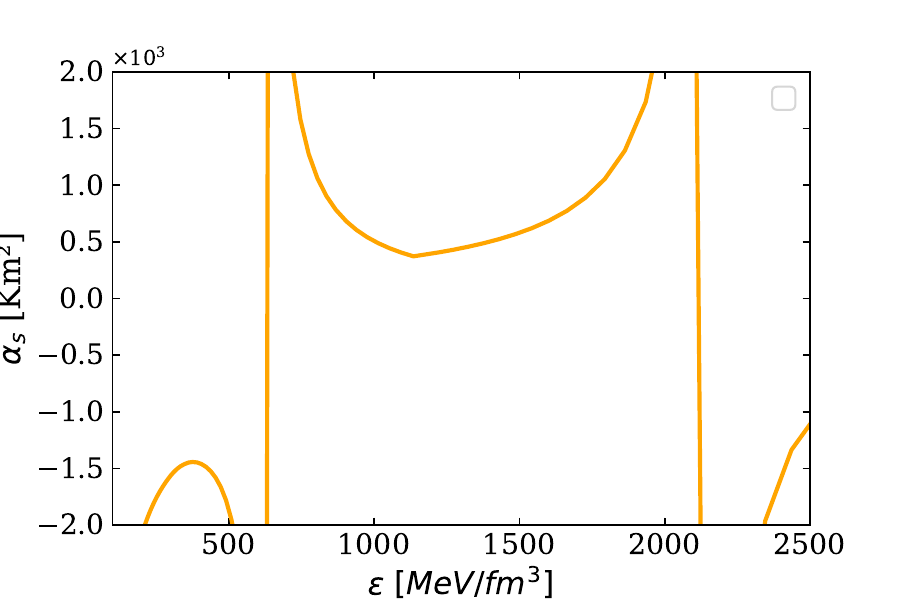}     \includegraphics[width=0.48\textwidth]{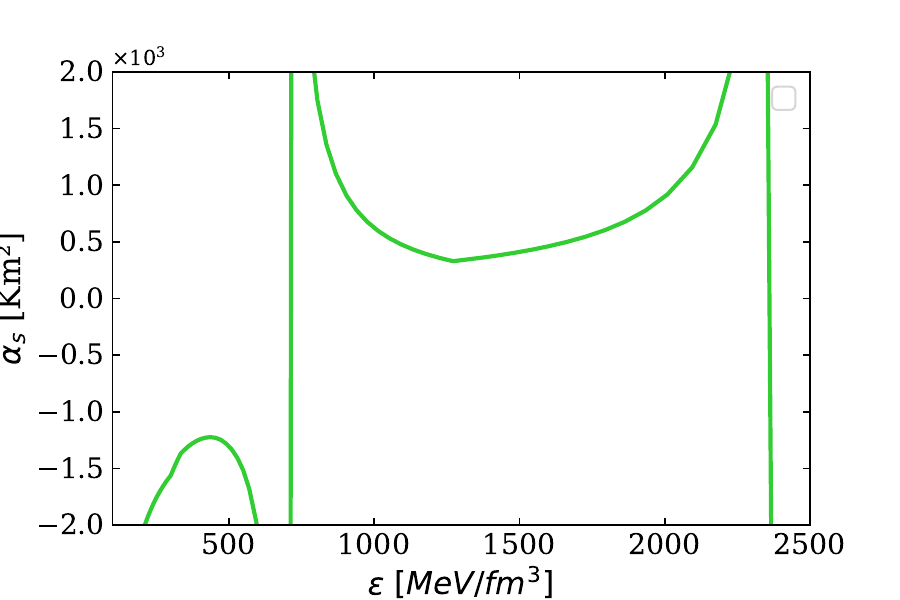}\\     \includegraphics[width=0.48\textwidth]{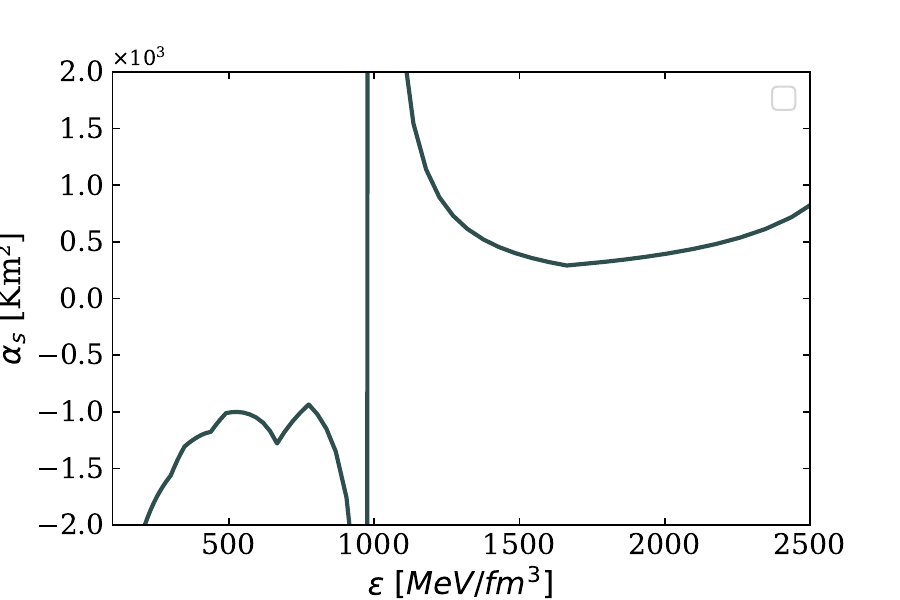}     \includegraphics[width=0.48\textwidth]{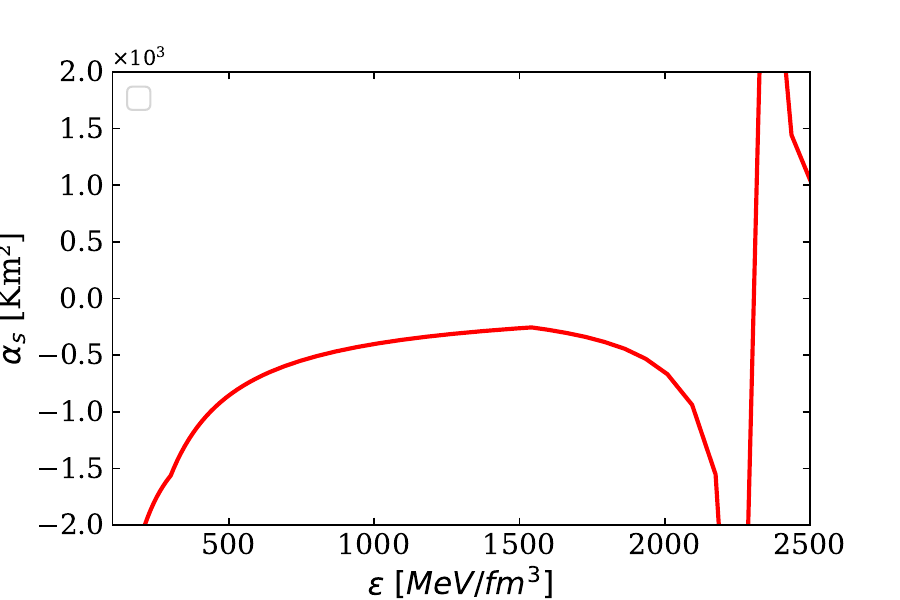}
    \caption{ \label{fig:maxminrange2}  Values of the parameter $\alpha_s$ for  which $M \rightarrow \infty$ for EoSs obtained at matching point $n_m=2n_s$ with $\chi$EFT EoSs of Fig. \ref{fig:EoSinterpol2}. {\bf Top:} Left, values of $\alpha_s$ for the stiffest (Blue) EoS; right,  The same values of$\alpha_s$ but at higher resolution. {\bf Second row:} Left, values of $\alpha_s$ for the Orange EoS. Right, Values of $\alpha_s$ for the Green EoS. {\bf Second row:}  Left, values of $\alpha_s$ for the Grey EoS (left); right, values of $\alpha_s$ for the softest (Red) EoS. These later figures show that the allowed values of  $\alpha_s$ for the energy density range $ 50 < \varepsilon < 1000$ MeV/fm$^3$  is $-2700 \lesssim \alpha \lesssim 550$ km$^2$, for the Blue EoS. For the Red EoS, $-1650 \lesssim\alpha \lesssim -250$ km$^2$ for energy densities $50 < \varepsilon < 2200$ MeV/fm$^3$. Note that taking into account another ranges of energy-density, $\alpha_s$ can overcome these values.}
   \end{figure}

The values of the singular parameter $\alpha_s$ for each EoS are displayed in Figure \ref{fig:maxminrange2}. As observed, there is a prominent peak at the sign change of $\varepsilon - 3P$. These graphs indicate that the stiffer the EoS, the wider the range of allowed $\alpha$ values.

Based on this analysis, we focus on the range $-1000 \leq \alpha \leq 1000$ for each EoS considered in this section.

\subsubsection{Constraining Starobinsky parameter}
\begin{figure}
    \centering
    \includegraphics[width=0.9\textwidth]{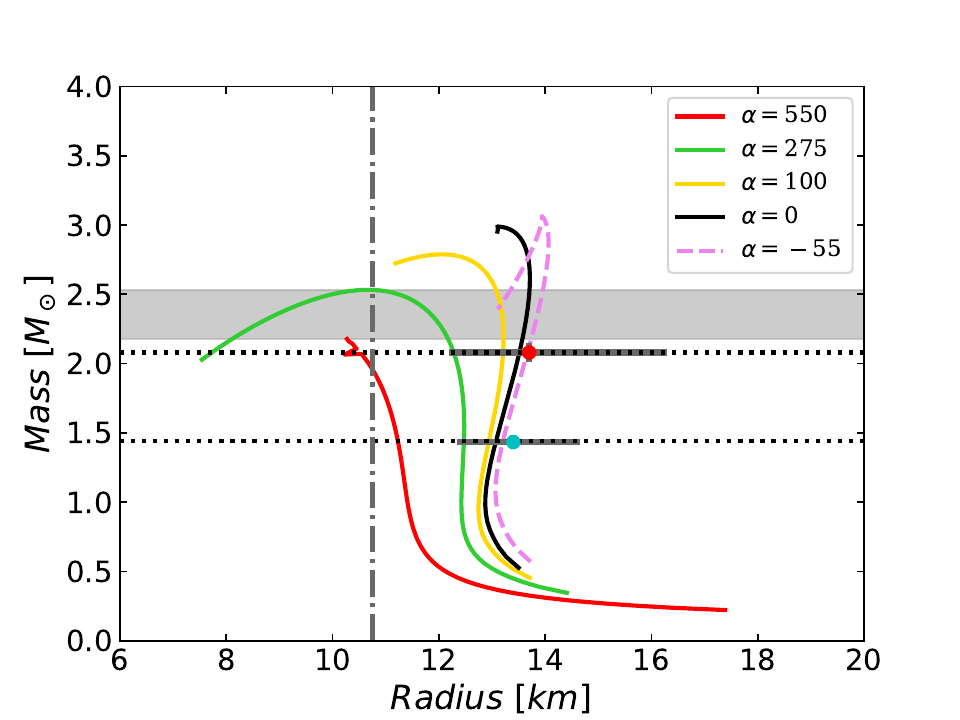}
         \includegraphics[width=0.48\textwidth]{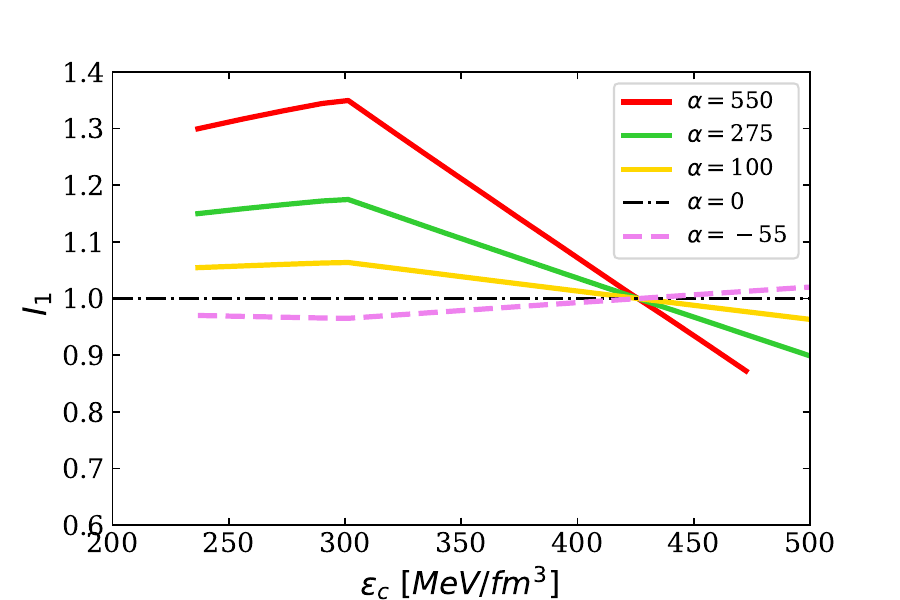}
     \includegraphics[width=0.48\textwidth]{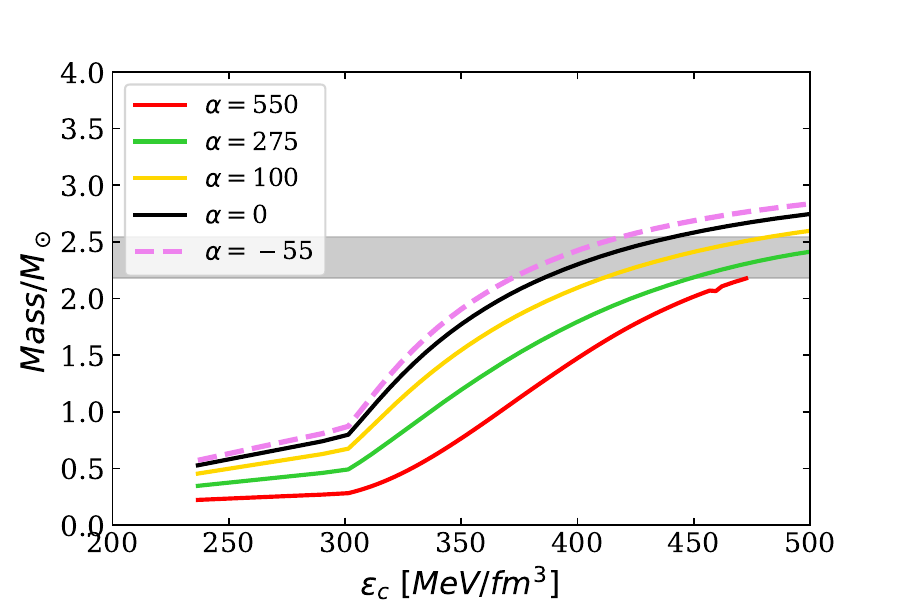}\\
     \includegraphics[width=0.48\textwidth]{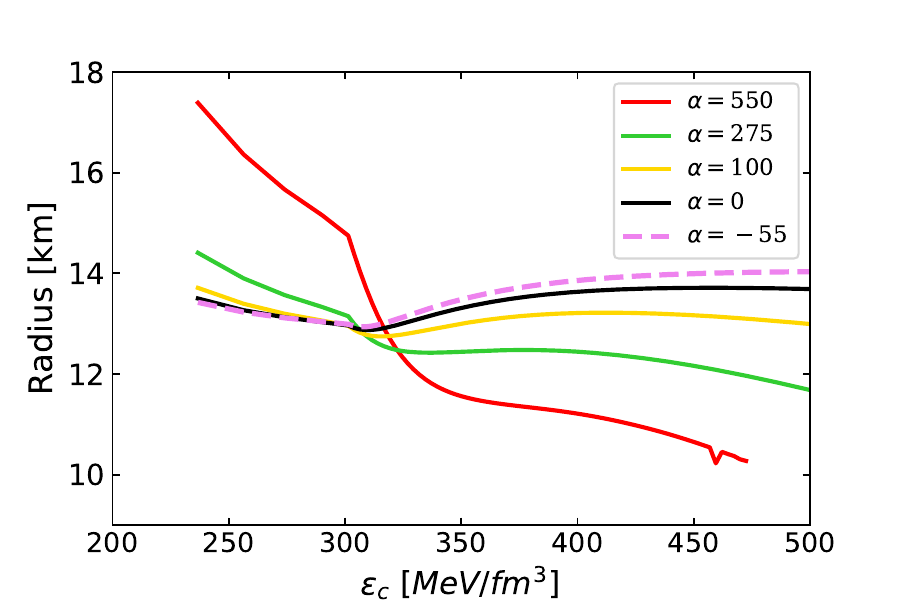}
     \includegraphics[width=0.48\textwidth]{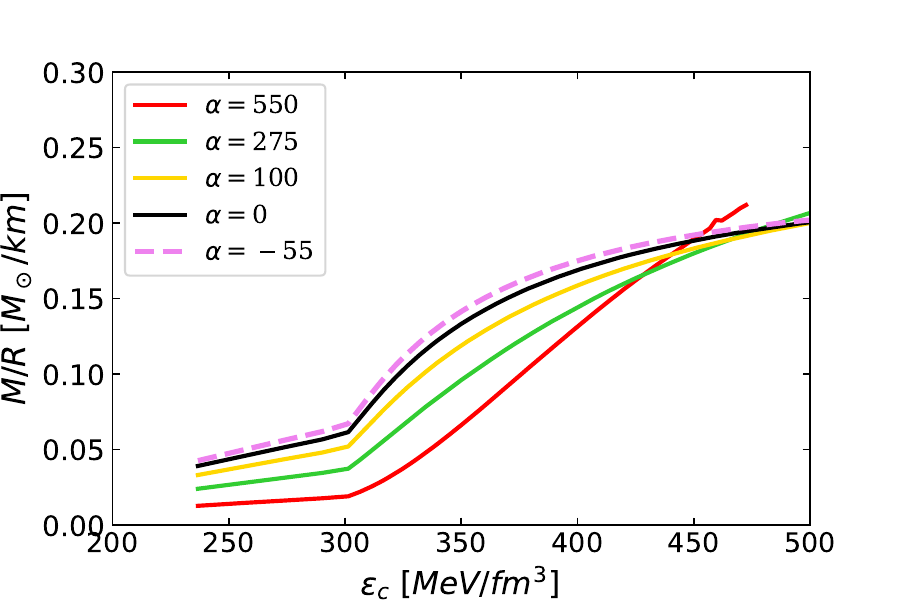}

    \caption{{\bf Top:} Mass-Radius diagrams in GR (black solid line) and ETG (coloured lines) for the maximum (Blue) EoS at the bottom of Fig. \ref{fig:EoSinterpol2}. The grayish shaded area refers to the  lower and  upper mass bound (2.17 and 2.52 M$_\odot$, respectively) and the vertical dash-dotted line indicates the lower radius bound (10.8 km). {\bf Second row:} $\mathcal{I}_1$ (left) and mass (right) against the  central energy density for the Blue EoS for different values of $\alpha$. {\bf Third row:} Radius and compactness against the central energy density for Blue EoS for different values of $\alpha$. }
    \label{fig:MvREoSmax2n}
\end{figure}

In this band,  we can observe in Fig. \ref{fig:MvREoSmax2n} how the stiffest EoS (Blue EoS) significantly reduces the maximum mass and radius for both gravity models compared to the interpolation at $n=n_s$. Additionally, the range of allowed $\alpha$ values narrows down from ($955-1030$) km$^2$ to ($275-550$) km$^2$.

Regarding the Red EoS (the softest), the range of $\alpha_s$ is very similar to the softest EoS in the $n_s$ band ($-250 <\alpha < 250$ km$^2$ for $500< \varepsilon <2000$ MeV/fm$^3$). However, unlike the previous case, the solutions of the TOV equations in GR yield a neutron star with a mass of $M=0.50 M_\odot$ at $\varepsilon=230$ MeV/fm$^3$. Therefore, we can search for allowed values of $\alpha$ at this energy density, where $\alpha_s > -1800$ km$^2$ (see Fig. \ref{fig:maxminrange2}, bottom right). Indeed, this EoS allows for $\alpha$ values on the order of $\sim -10^3$ for central energy densities between $230< \varepsilon < 296$ MeV/fm$^3$. It is worth mentioning that for $\alpha=-1510$ km$^2$, the Red EoS reaches a maximum mass of $M=2.41 M_\odot$ at a radius of $R=13.36$ km for $\varepsilon=296$ MeV/fm$^3$.

\begin{figure}
      \centering
      \includegraphics[width=0.80\textwidth]{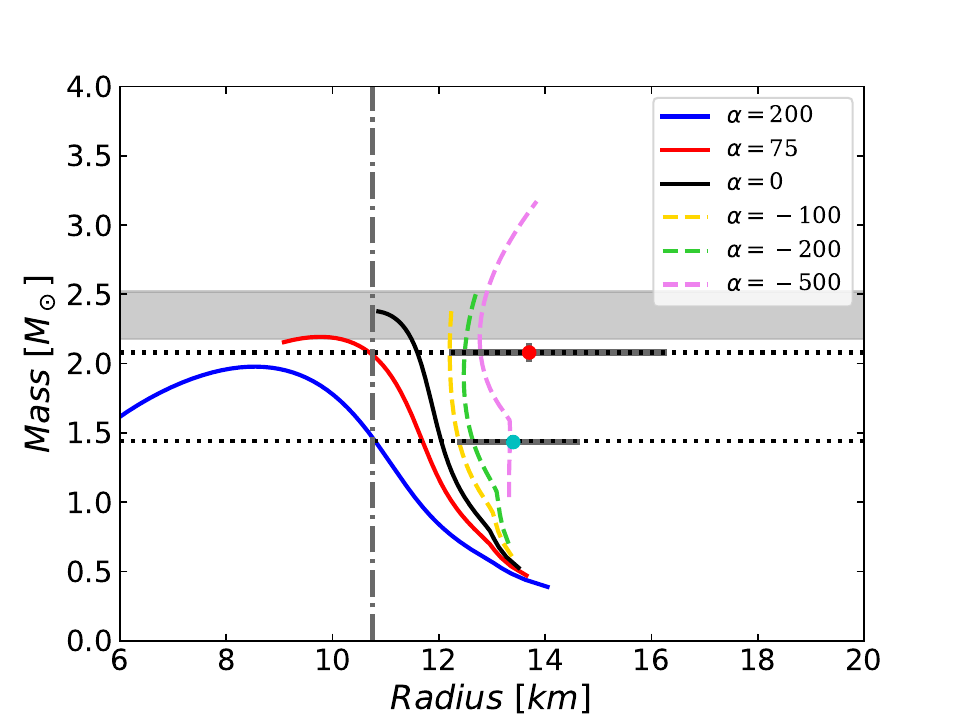}      \includegraphics[width=0.45\textwidth]{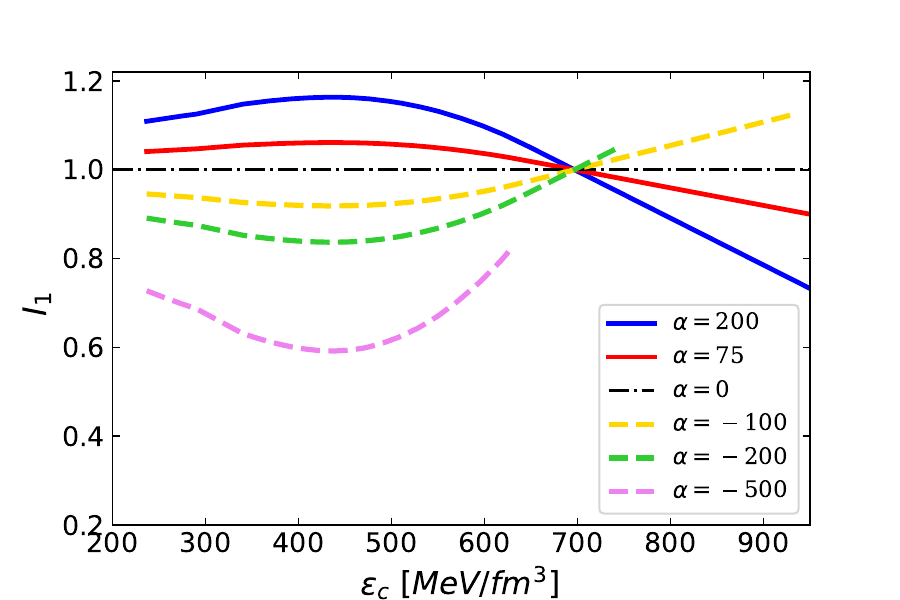}
     \includegraphics[width=0.45\textwidth]{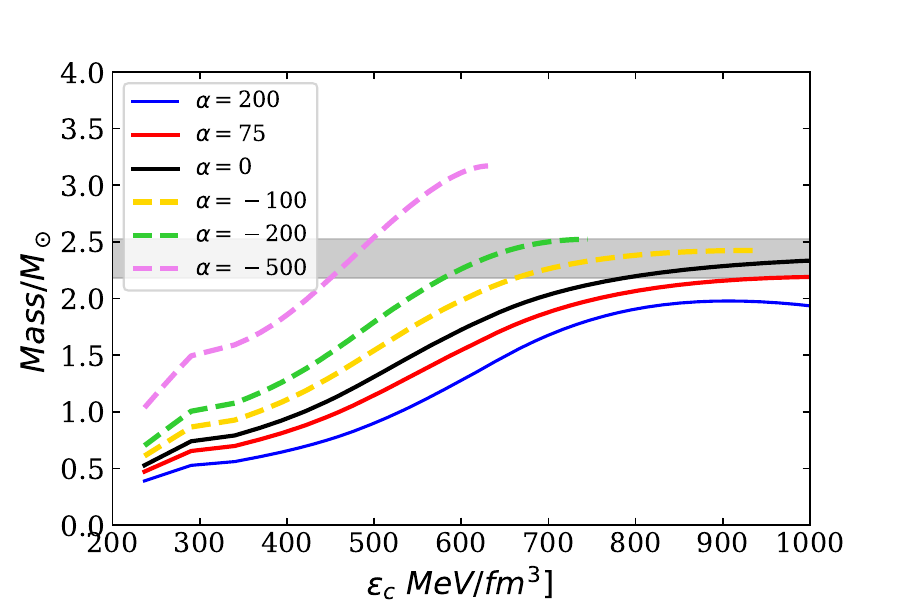}\\
     \includegraphics[width=0.45\textwidth]{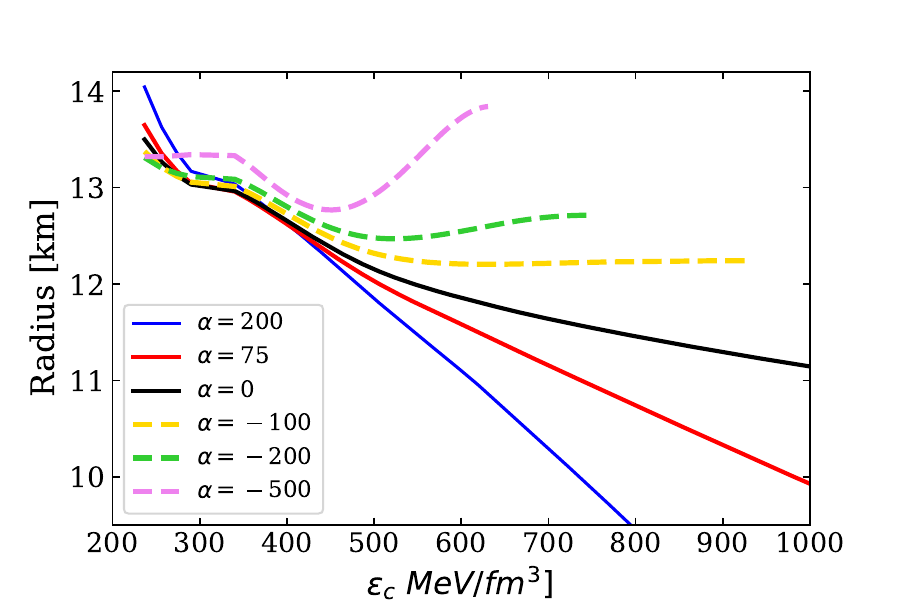}
     \includegraphics[width=0.45\textwidth]{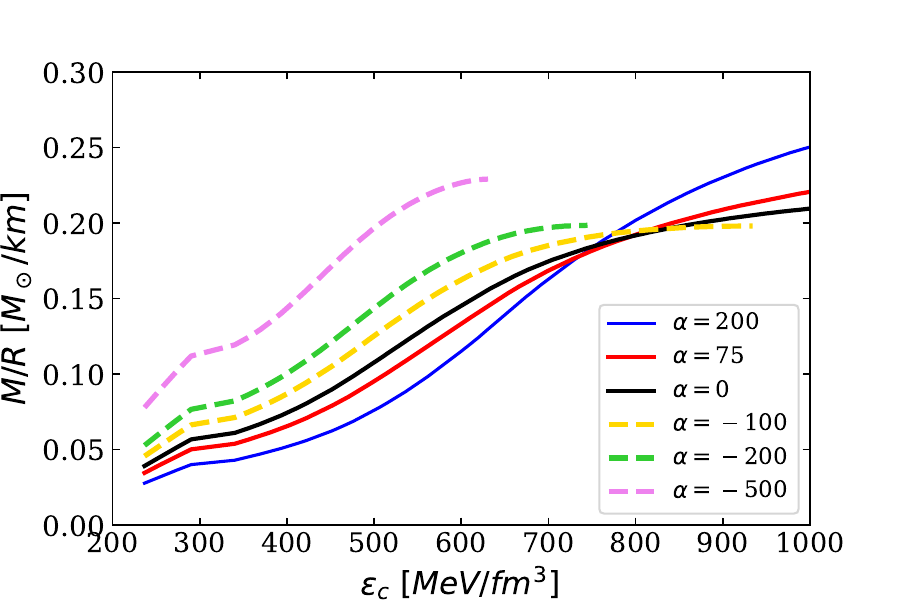}
     \caption{{\bf Top:} Mass-Radius diagrams in GR (black solid line) and ETG (coloured lines) for the next stiffest (Green) EoS of \ref{fig:EoSinterpol2}. The grayish shaded area refers to the lower and  upper mass bound (2.17 and 2.52 M$_\odot$, respectively) and the vertical dash-dotted line indicates the lower radius bound (10.8 km). {\bf Second row:} $\mathcal{I}_1$ (left) and mass (right) against the central energy density for the Green EoS \ref{fig:EoSinterpol2}  for different values of $\alpha$. {\bf Third row:} Radius and compactness against the central energy density for the Green EoS than above for different values of $\alpha$. }
      \label{fig:MvREoSintermPT12n}
  \end{figure}

\begin{figure}
      \centering
      \includegraphics[width=0.80\textwidth]{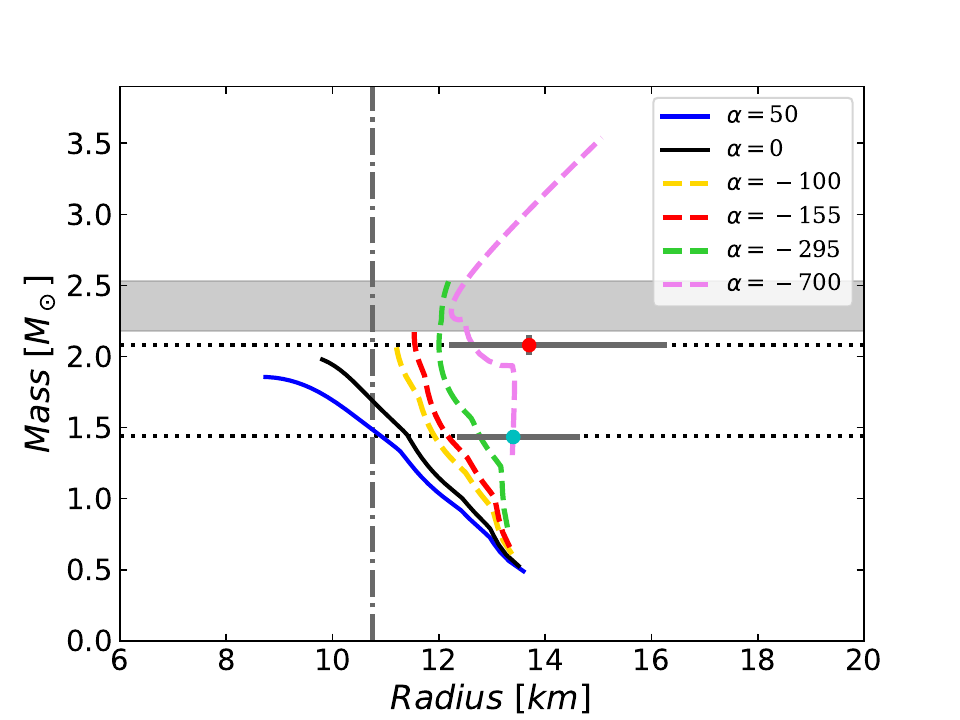}
      \includegraphics[width=0.45\textwidth]{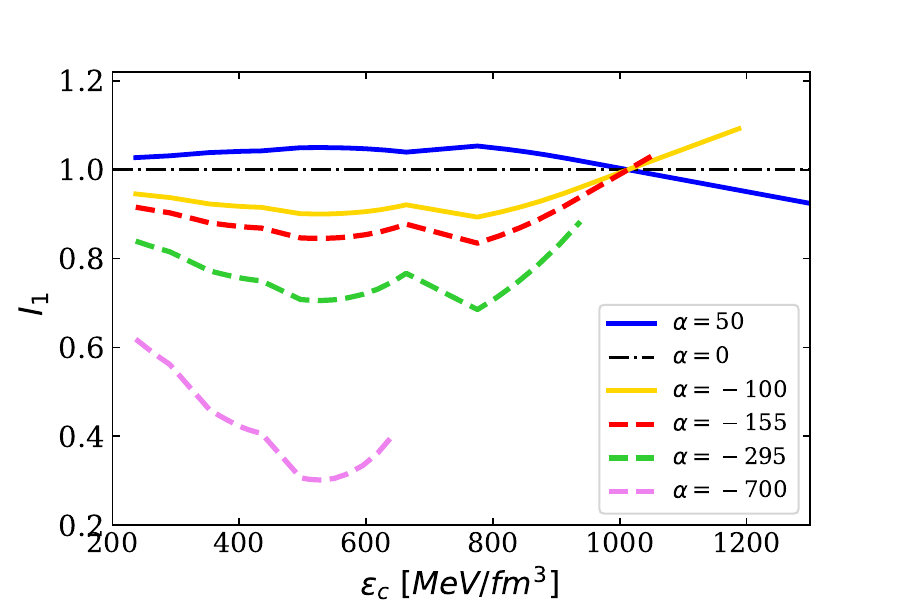}
     \includegraphics[width=0.45\textwidth]{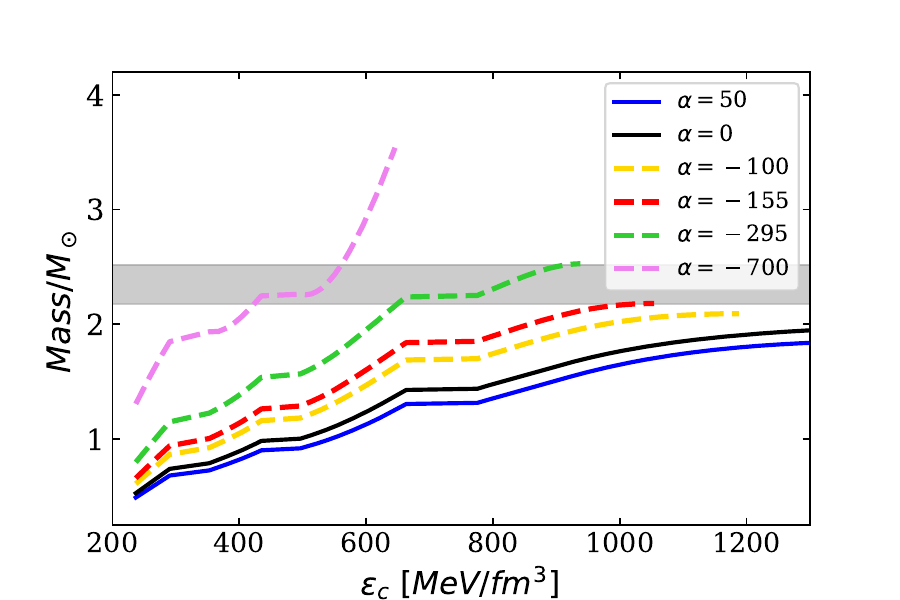}\\
     \includegraphics[width=0.45\textwidth]{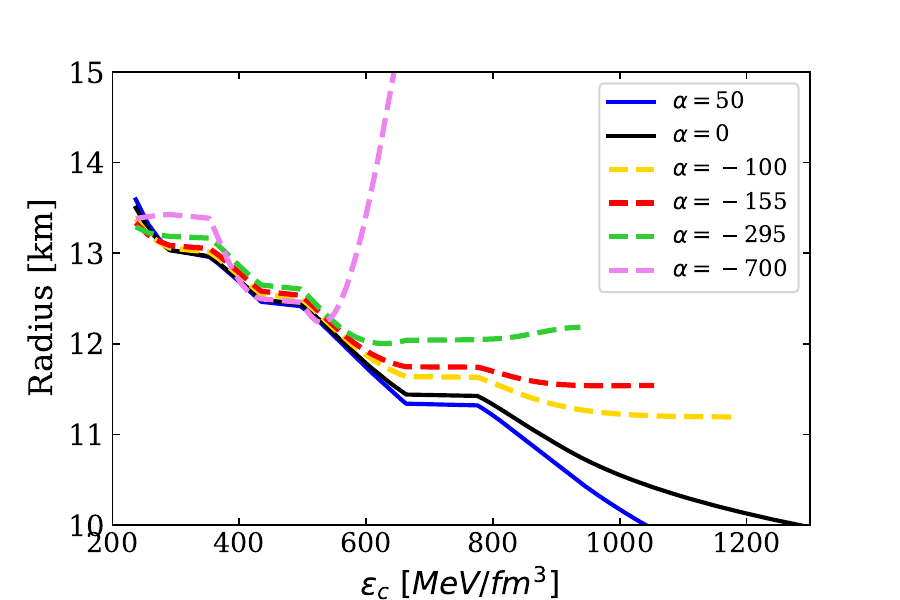}
     \includegraphics[width=0.45\textwidth]{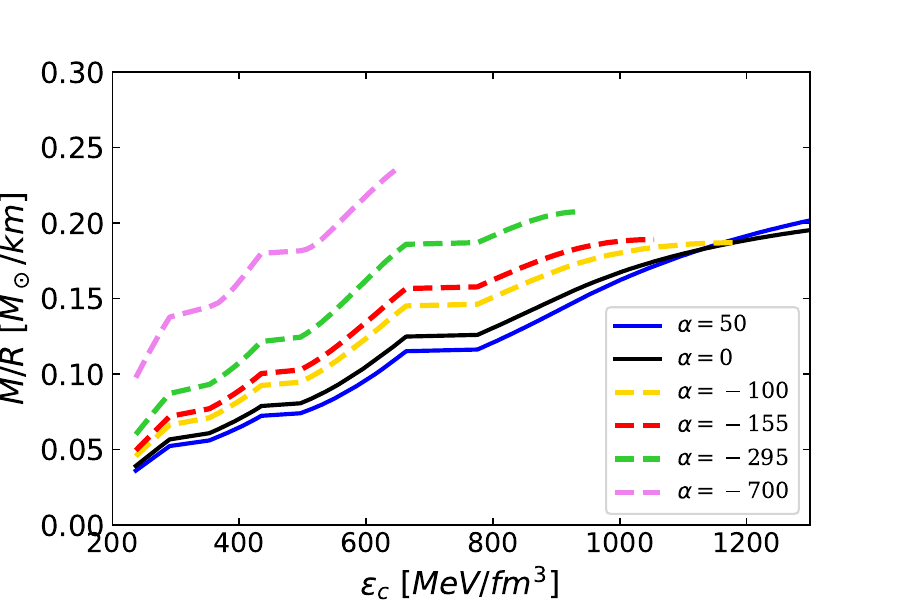}      
      \caption{Mass-Radius diagrams in GR (black solid line) and ETG (coloured lines) for the Grey EoS of \ref{fig:EoSinterpol2}. The grayish shaded area refers to the  lower and  upper mass bound (2.17 and 2.52 $M_\odot$, respectively) and the vertical dash-dotted line indicates the lower radius bound (10.8 km). {\bf Second row:} $\mathcal{I}_1$ (left) and mass (right) against the  central energy density for the Grey EoS  for different values of $\alpha$. {\bf Third row:} Radius and compactness against  the central energy density for the Grey EoS  for different values of $\alpha$. }
      \label{fig:MvREoSintermPT32n}
  \end{figure}

Another noteworthy aspect is that for EoSs with a first-order phase transition, such as the Green and Grey1st EoSs shown in Fig. \ref{fig:EoSinterpol2}, Palatini gravity alters the shape of the mass-radius curves compared to the GR plots, as depicted in Fig. \ref{fig:MvREoSintermPT12n} and \ref{fig:MvREoSintermPT32n}.

  \begin{figure}
     \centering
     \includegraphics[width=0.90\textwidth]{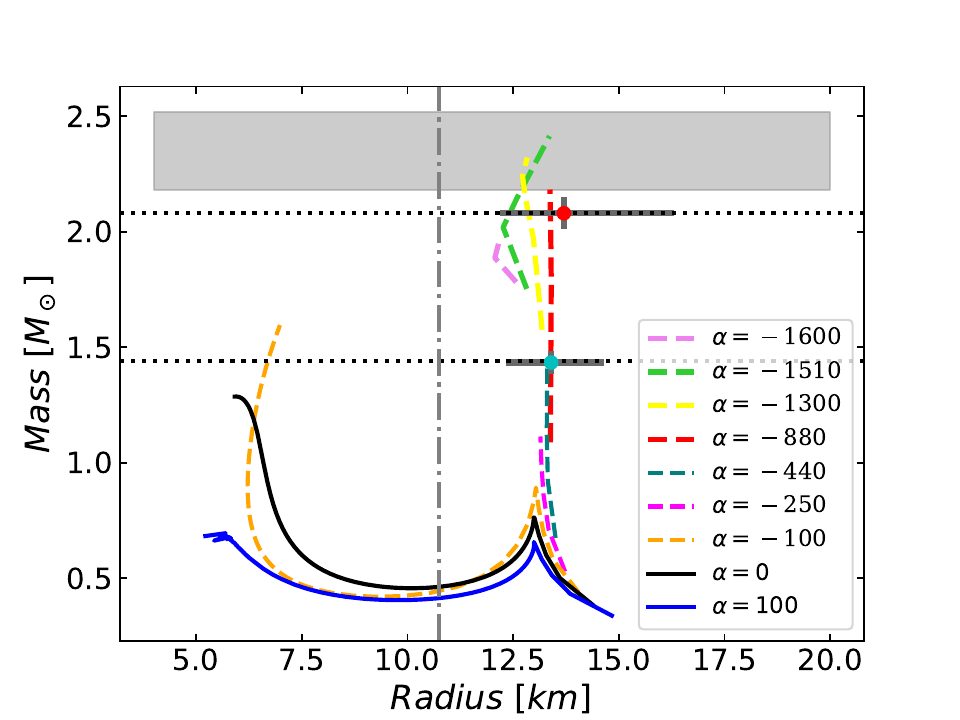}
      \caption{Mass-Radius diagrams in GR (black solid line) and ETG (coloured lines) for the minimum (Red) EoS of \ref{fig:EoSinterpol2}. The grayish shaded area refers to the  lower and  upper mass bound (2.17 and 2.52 $M_\odot$, respectively) and the vertical dash-dotted line indicates the  lower radius bound (10.8 km).}
      \label{fig:MvREoSmin2n}
 \end{figure}

In summary, the results obtained by interpolating with the chiral EoSs at $n=2n_s$ exhibit similar characteristics to those obtained by interpolating at the saturation density, as shown in Table \ref{tab:alphabounds}, where both results are collected.

  \begin{figure}
      \centering
      \includegraphics[width=0.80\textwidth]{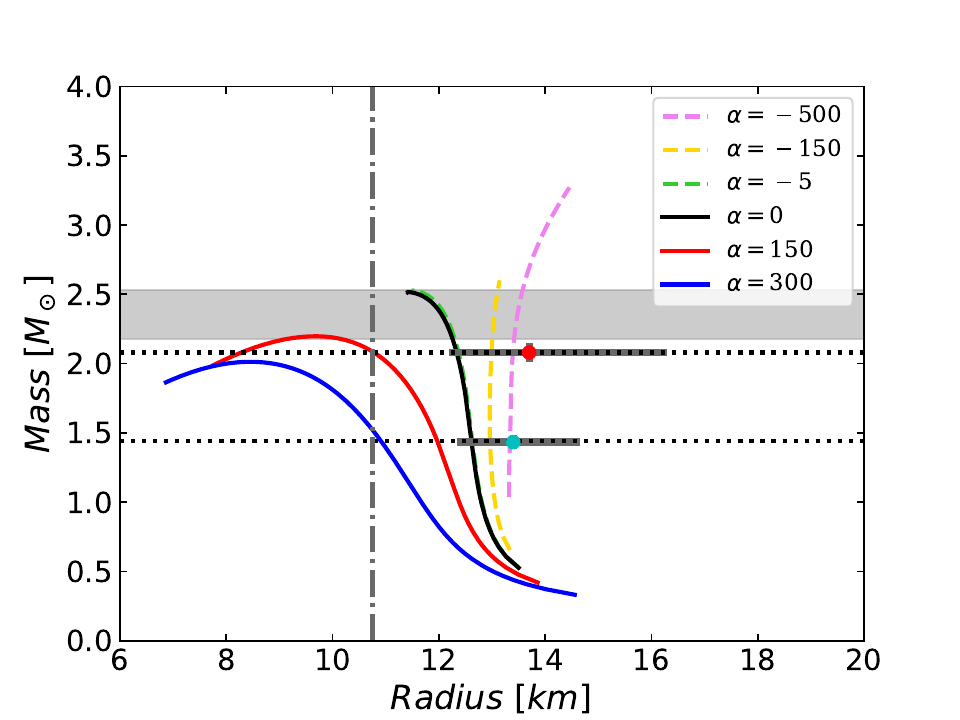}
      \includegraphics[width=0.45\textwidth]{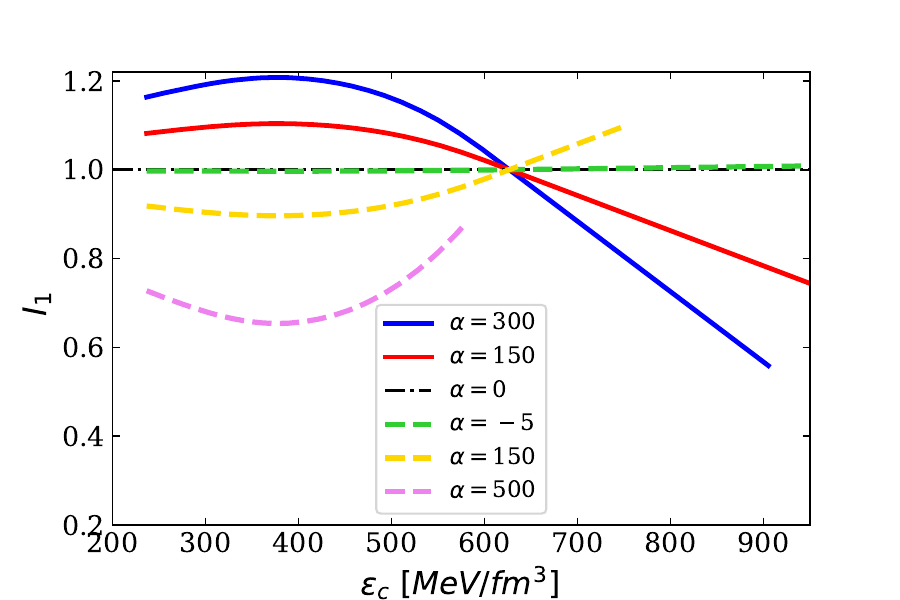}
     \includegraphics[width=0.45\textwidth]{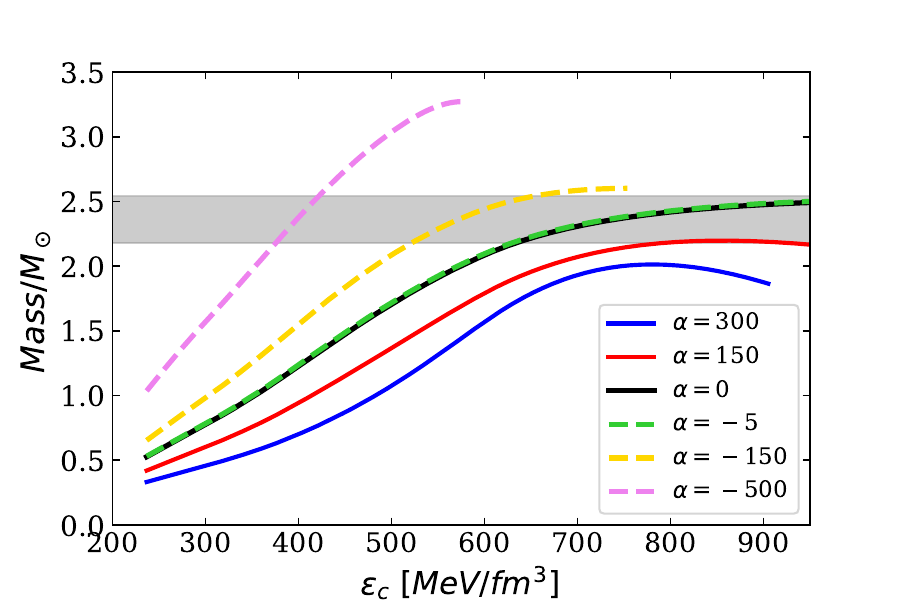}\\
     \includegraphics[width=0.45\textwidth]{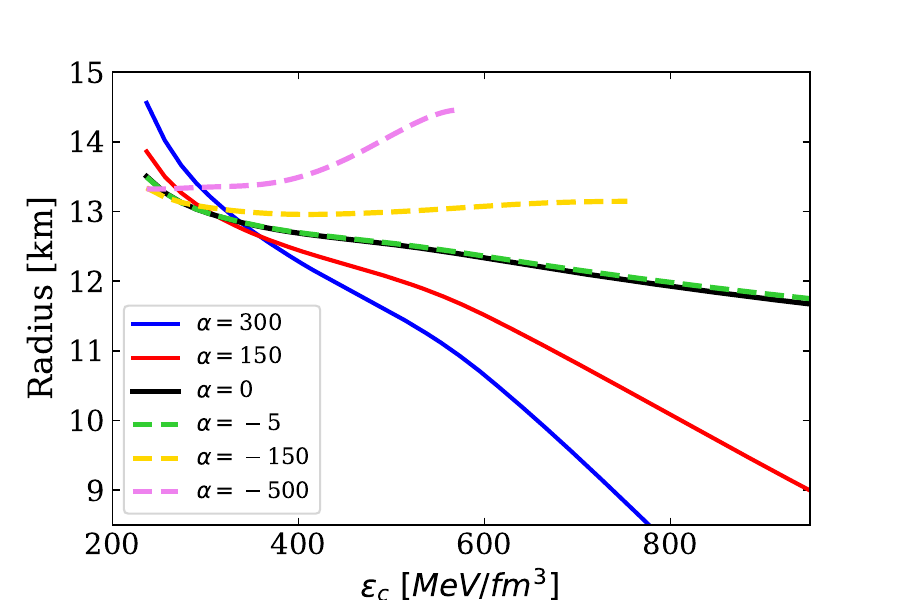}
     \includegraphics[width=0.45\textwidth]{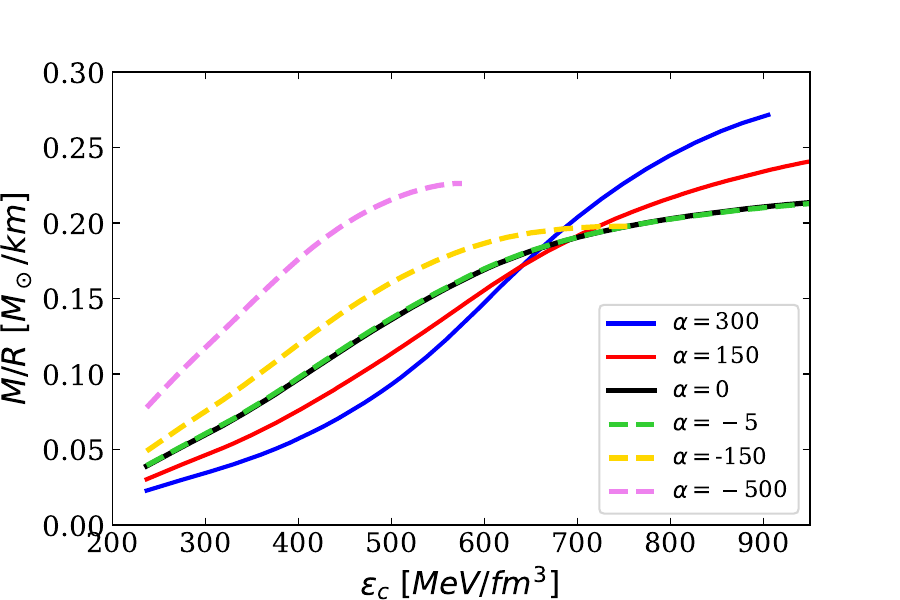}
      \caption{Mass-Radius diagrams in GR (black solid line) and ETG (coloured lines) for the next stiffest (Orange) EoS of \ref{fig:EoSinterpol2}. The grayish shaded area refers to the lower and upper mass bound (2.17 and 2.52 $M_\odot$, respectively) and the vertical dash-dotted line indicates the lower radius bound (10.8 km).}
      \label{fig:MvREoSinterm12n}
  \end{figure}

\begin{table}
    \centering
    \begin{tabular}{c c c}
    \hline
      EoS & $\alpha$ ($n$ = $n_s$) (km$^2$) & $\alpha$ ($n$ = 2$n_s$) (km$^2$) \\ \hline \hline
        Blue & (955, 1030) & (275, 550) \\
        Orange & (250, 625 ) & (-5, 150)   \\
        Green & (-175, 60) & (-200, 75) \\
        Dark green & (-320, -210) & (-300, -150)  \\
        Red & - & (-1510,  -880) 
    \end{tabular}
    \caption{Constraints on $\alpha$ for EoS of Fig. \ref{fig:EoSinterpol2},  by setting bounds from astrophysical observations}
    \label{tab:alphabounds}
\end{table}

\subsection{Brief comment on phase transitions}

\begin{figure}
      \centering      \includegraphics[width=0.45\textwidth]{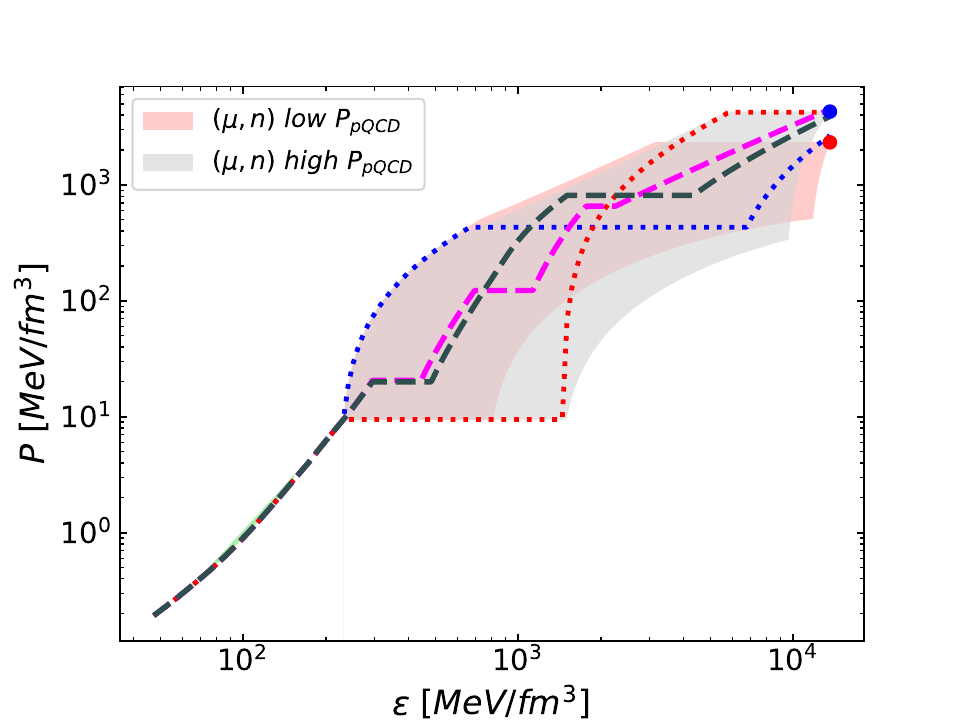}      \includegraphics[width=0.45\textwidth]{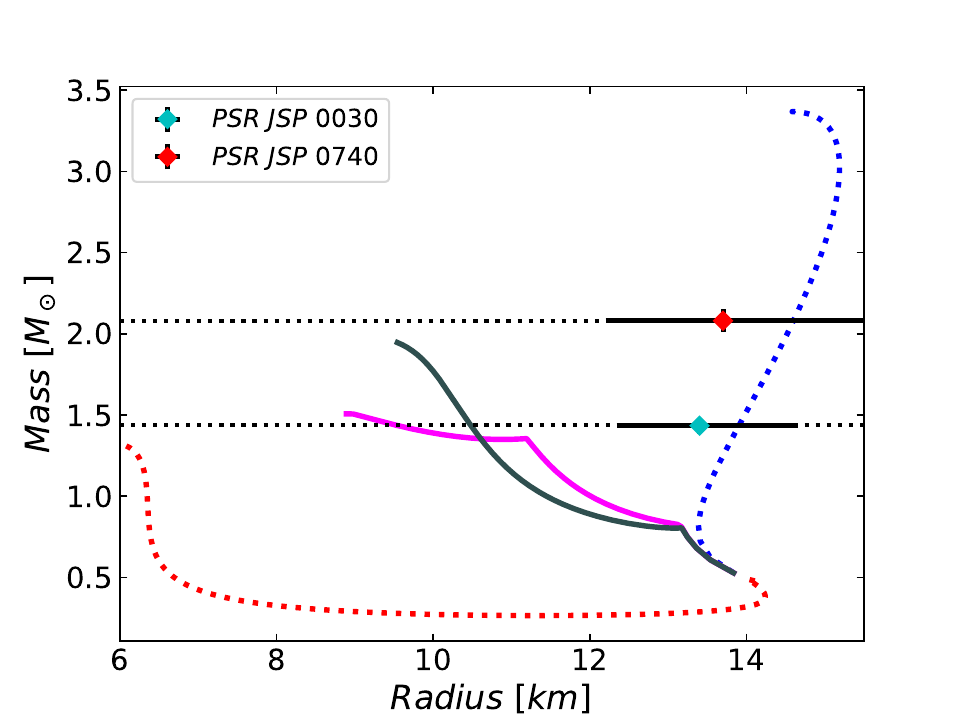}\\     \caption{{\bf Top:} The stiffest EoS (Blue), the softest (Red), and the two intermediate ones (Grey and Magenta) at the interpolation region obtained at $n=1.5 n_s$ (left) and corresponding M-R diagrams in GR (right). We focus on two middle ones (Grey and Magenta) with one and three phase transitions, respectively. 
      }
      \label{fig:EoS1_5n}
  \end{figure}

\begin{figure}
      \centering      \includegraphics[width=0.80\textwidth]{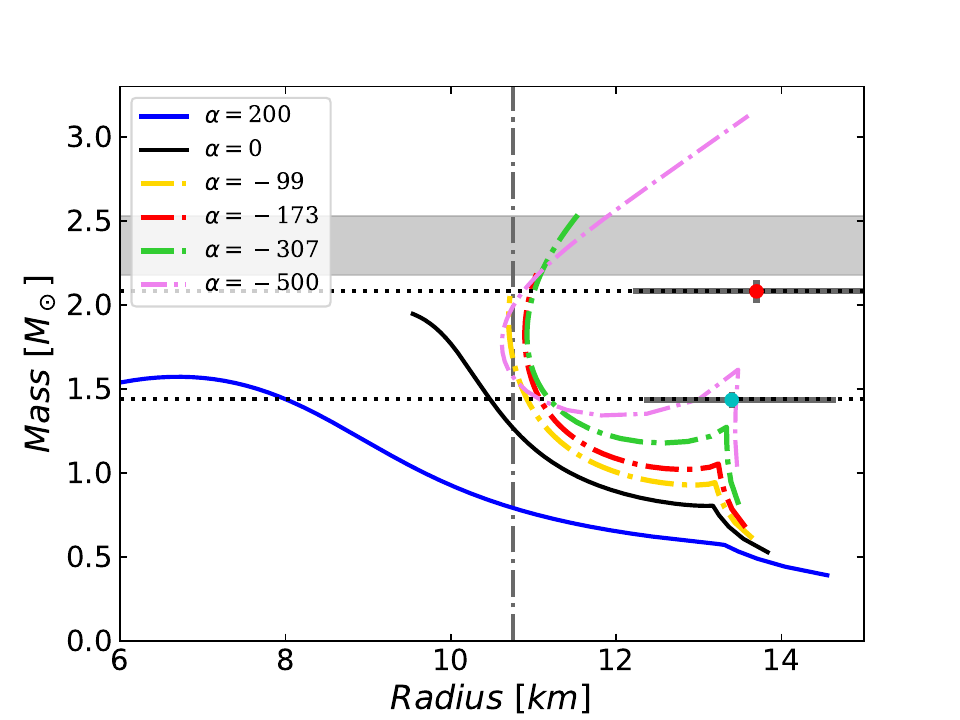}      \includegraphics[width=0.45\textwidth]{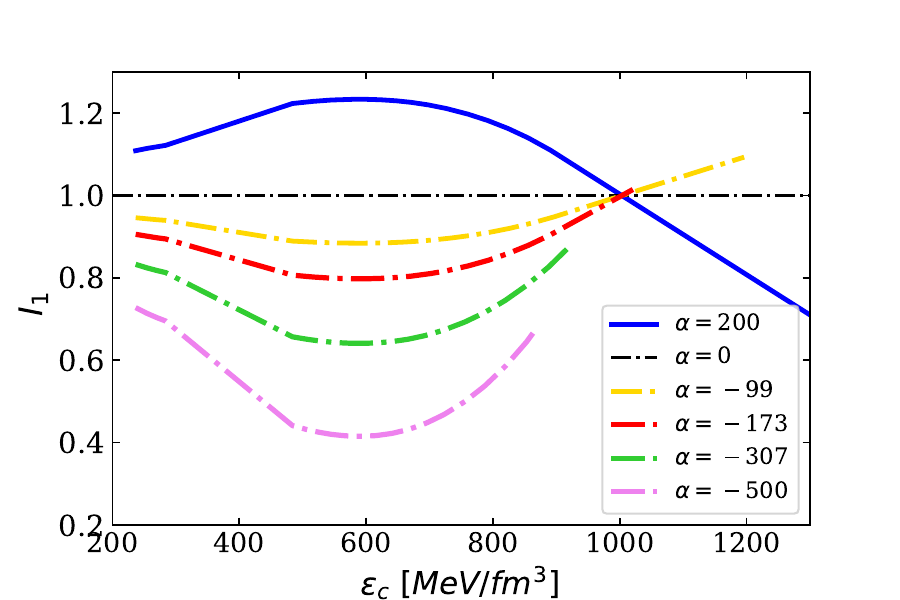}     \includegraphics[width=0.45\textwidth]{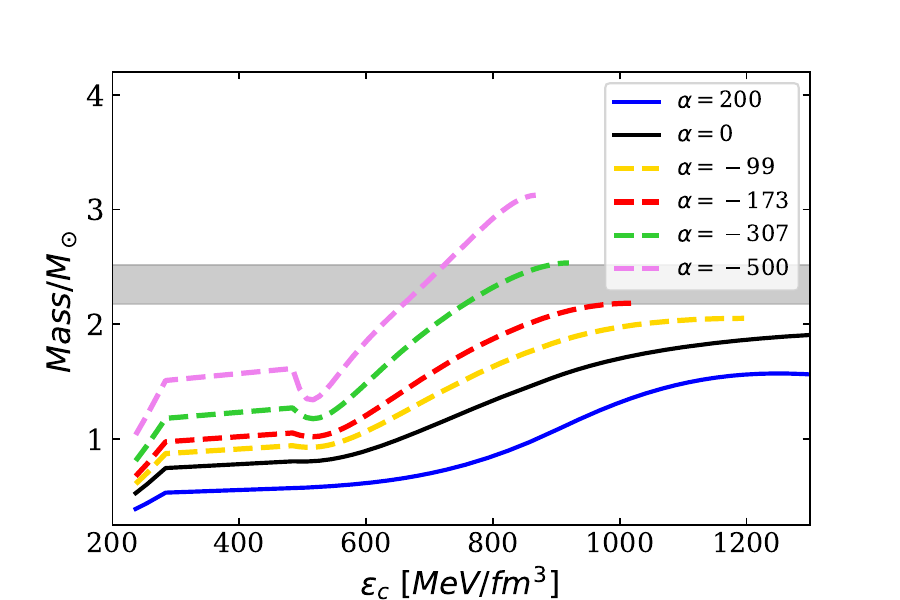}\\
     \includegraphics[width=0.45\textwidth]{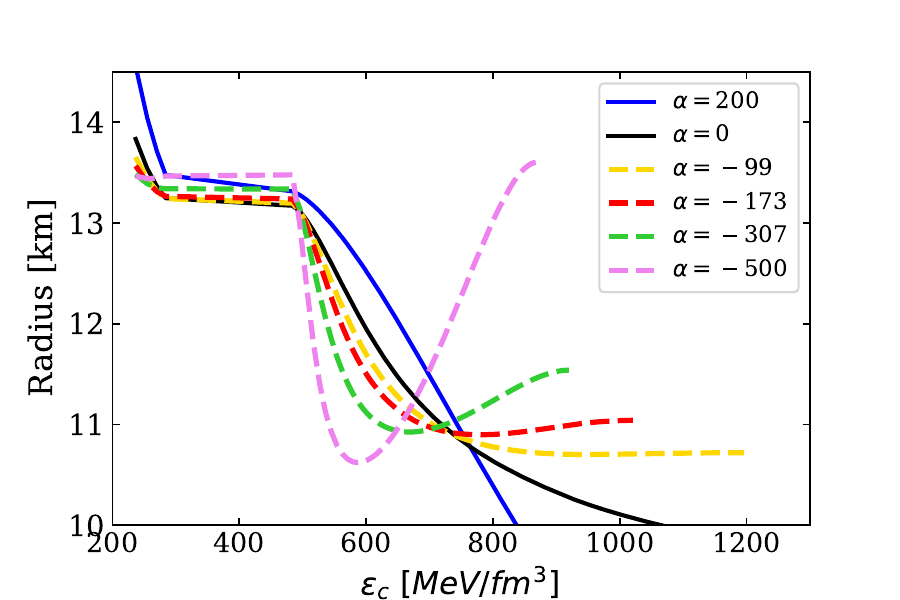}
     \includegraphics[width=0.45\textwidth]{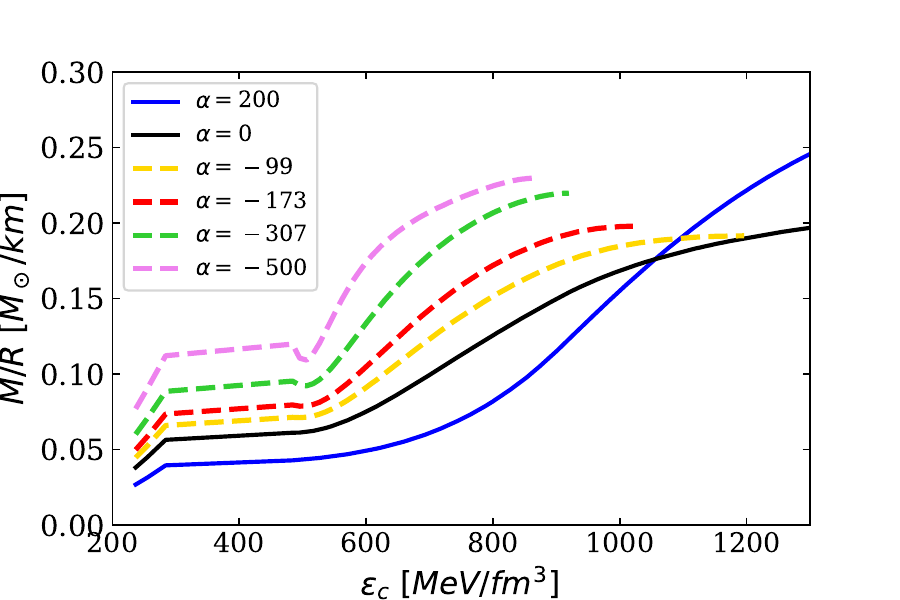}
     \caption{{\bf Top:} Mass-Radius diagrams in GR (black solid line) and ETG (coloured lines) for the Yellow EoS  of Fig. \ref{fig:EoS1_5n}. The grayish shaded area refers to the  lower and  upper mass bound (2.17 and 2.52 $M_\odot$, respectively) and the vertical dash-dotted line indicates the lower radius bound (10.8 km). {\bf Second row:} $\mathcal{I}_1$ (left) and Mass (right) against the  central energy density for the Yellow EoS \ref{fig:EoS1_5n}  for different values of $\alpha$. {\bf Third row:} Radius and compactness against  the central energy density for the Yellow EoS for different values of $\alpha$. }
      \label{fig:MvREoSintermlow0bis}
  \end{figure}

\begin{figure}
      \centering      \includegraphics[width=0.80\textwidth]{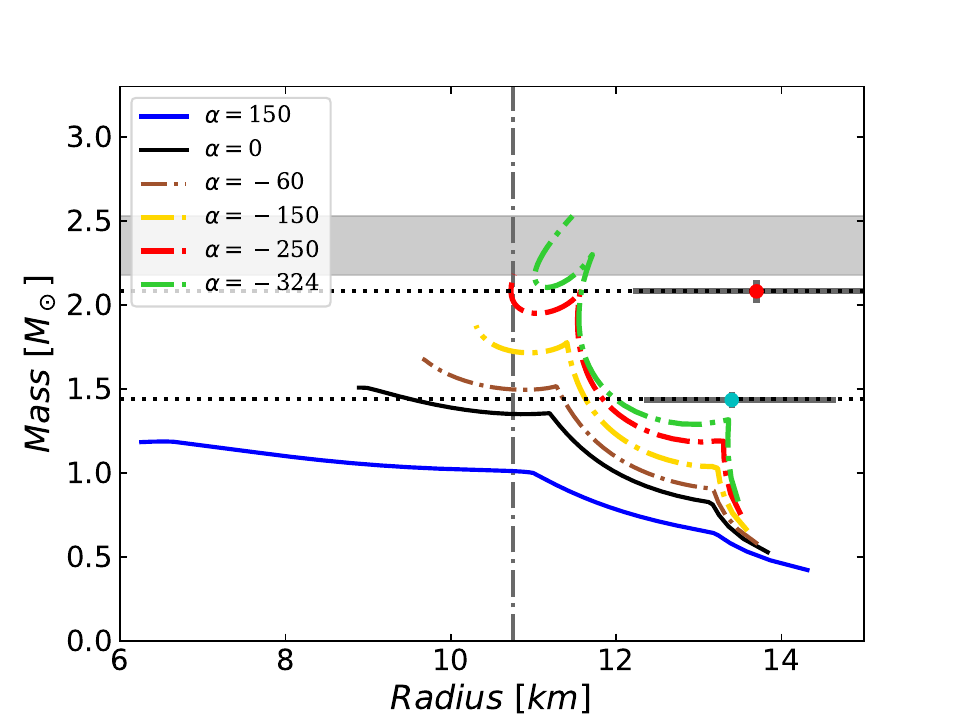}      \includegraphics[width=0.45\textwidth]{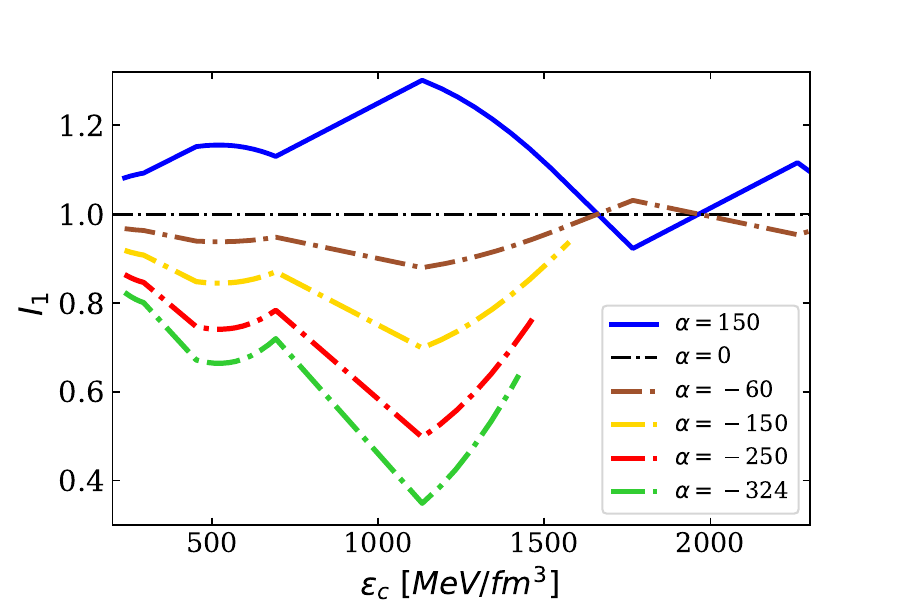}     \includegraphics[width=0.45\textwidth]{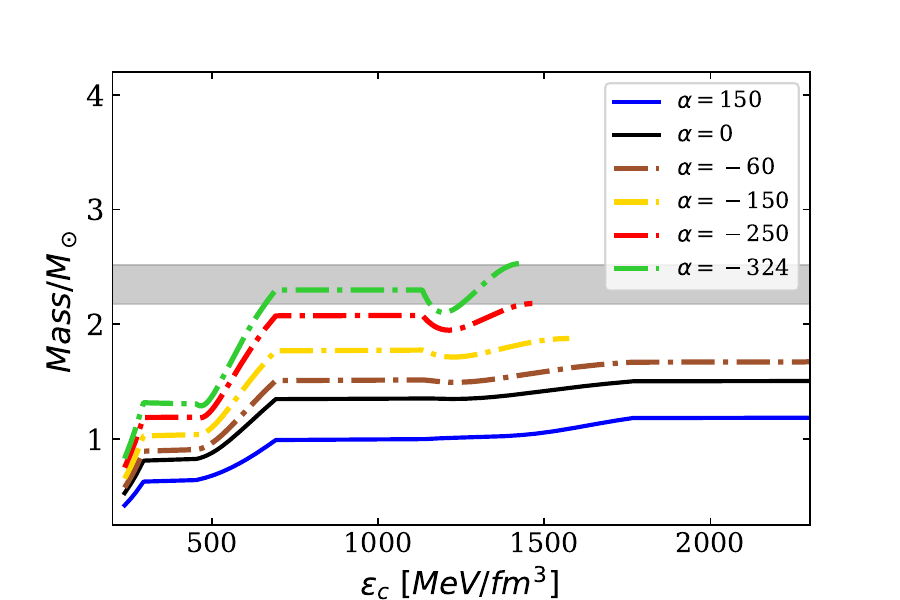}\\
     \includegraphics[width=0.45\textwidth]{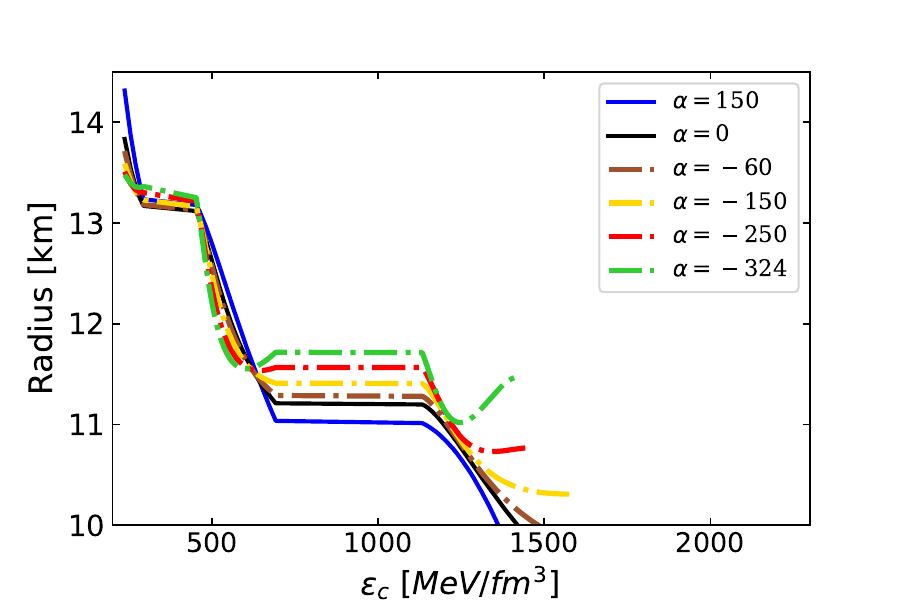}
     \includegraphics[width=0.45\textwidth]{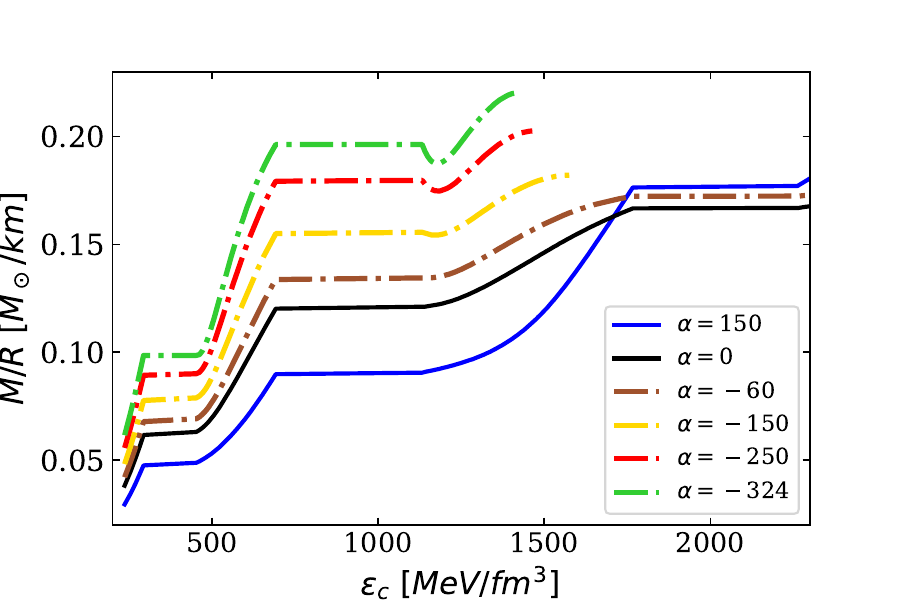}
     \caption{{\bf Top:} Mass-Radius diagrams in GR (black solid line) and ETG (coloured lines) for the Magenta EoS  of Fig. \ref{fig:EoS1_5n}. The grayish shaded area refers to the  lower and  upper mass bound (2.17 and 2.52 $M_\odot$, respectively) and the vertical dash-dotted line indicates the lower radius bound (10.8 km). {\bf Second row:} $\mathcal{I}_1$ (left) and mass (right) against the  central energy density for the Magenta EoS for different values of $\alpha$. {\bf Third row:} Radius and compactness against the central energy density for the Magenta EoS for different values of $\alpha$. }
      \label{fig:MvREoSinterm3PT20}.
  \end{figure}

Let us now briefly consider the interpolation at $n=1.5n_s$. The reason for examining this case is that it lies between the other two interpolations at $n=n_s$ and $n=2n_s$, allowing us to confirm and compare the results obtained in those bands, particularly regarding phase transitions. Thus, we will focus on two equations of state as shown in Fig. \ref{fig:EoS1_5n}: the Grey EoS with one phase transition, and the Magenta EoS with three phase transitions. The corresponding results are presented in Fig. \ref{fig:MvREoSintermlow0bis} and Fig. \ref{fig:MvREoSinterm3PT20}, respectively.

The Grey EoS in Fig. \ref{fig:EoS1_5n} exhibits a long phase transition at $P=20$ MeV/fm$^{-3}$ within the energy density range $\varepsilon=295-483$ MeV/fm$^{-3}$ ((1.80 - 3.01) $n_s$). However, it is worth noting that the jump in energy density slightly exceeds the Seidov limit defined in \cite{Seidov:1971sv}:
\begin{equation}\label{Seidovsjump}
\Delta \varepsilon_{\text{crit}} := \varepsilon_E-\varepsilon_H = \varepsilon_H \left( \frac{1}{2} +\frac{3}{2}\frac{P_H}{\varepsilon_H} \right),
\end{equation}
which determines the maximum energy density a neutron star core can sustain before collapsing into a black hole (in the perturbative, small-core approximation). It is important to note that this property is dependent on GR. In \cite{Lope-Oter:2021mjp} a definition of latent heat is proposed to characterize the first-order phase transition based on purely hadronic calculations. Since the Seidov's criterion cannot be used with modified gravity theories, the limit of \cite{Lope-Oter:2021mjp} may be the relevant one.

The Magenta EoS in Fig. \ref{fig:EoS1_5n} exhibits three long phase transitions at $P=20.6$ ($295\leq \varepsilon \leq 444$ MeV/fm$^3$), $122.8$ ($694\leq \varepsilon \leq1131$ MeV/fm$^3$), and $656.5$ MeV/fm$^3$ ($1769\leq \varepsilon\leq 2262$ MeV/fm$^3$). However, none of these phase transitions exceed the Seidov's limit in GR.

\begin{table}
    \centering
    \begin{tabular}{ccccc}
    \hline
      EoS & $M_{max}/M_{\odot}$ & R$_{M_{max}}$ & \ $\varepsilon_c$ (MeV/fm$^3$) &$\alpha$ (km$^{2}$) \\ 
      \hline \hline
        Max (Blue) & 3.37 & 14.56 & 667 & 600$< \alpha<$ 675  \\
         Magenta & 1.51 & 8.89 & 2383  & -324$< \alpha<$ -250  \\
    Grey & 1.95 & 9.54 & 1448  & -173$< \alpha<$ -307 \\
        Min (Red) & 1.33 & 5.93 & 4347& --
    \end{tabular}
    \caption{Values of the maximum mass and corresponding radius and central energy density in GR for the EoSs of Fig. \ref{fig:EoS1_5n}, interpolating at $n=1.5 n_s$. We have also provided bounds on the parameter $\alpha$ of Palatini gravity with respect to the given EoSs.}
    \label{tab:GR3}
\end{table}

When $\Delta \varepsilon < \Delta \varepsilon_{\text{crit}}$, a stable connected hybrid branch continues from the hadronic branch. However, if $\Delta \varepsilon > \Delta \varepsilon_{\text{crit}}$, there is no stable connected branch. Instead, a "third family" of neutron stars emerges, known as the disconnected branch \cite{Alford:2015gna}. The fate of these unstable configurations has not been fully explored, and it could have implications for the formation and existence of twin stars \cite{Espino:2021adh}—neutron stars with the same mass but different radii. In this case, the phase transition induces a local destabilization in the neutron star sequence within a small range of central energy densities. However, the sequence stabilizes again due to the stiffening of the equation of state.

The results obtained for GR for the considered EoSs in this density branch are shown on the right of Fig. \ref{fig:EoS1_5n}. In contrast, Fig. \ref{fig:MvREoSintermlow0bis} and \ref{fig:MvREoSinterm3PT20} demonstrate that for both EoSs, the typical mass-radius diagram of a twin-star changes its concavity as the absolute value of negative $\alpha$ increases. Moreover, it can be observed that the instability also increases with decreasing $\alpha$. This instability is more pronounced in ETG compared to GR when considering larger negative values of $\alpha$ (e.g., $\alpha=-500$ km$^2$) for masses and radii outside the normal range. The remaining results for ETG for different values of $\alpha$ are summarized in Table \ref{tab:GR3}.

Similarly, it can be observed that the radius decreases as the negative value of $\alpha$ increases, remaining approximately constant for $\alpha \approx 100$ and increasing for smaller values of $\alpha$.

\section{Discussion and conclusions}

In this study, we aimed to use gravity-independent equations of state to constrain a specific gravity model, that is, Palatini $f(\hat R)$ gravity. These equations of state were constructed based on earlier work \cite{LopeOter:2019pcq}, which provided equations for matter in neutron stars without relying on observational data or GR.

The construction of these equations involved incorporating results from NN and 3N chiral potentials for low-density scenarios, perturbative QCD for high-density regimes, and constraints from monotonicity and causality for intermediate densities (referred to as the "interpolation regime"). This approach aimed to maintain model independence while adhering to fundamental physics principles.

Most equations of state used to describe neutron star interiors depend on astrophysical observations and GR. However, to study neutron stars within alternative gravity theories, it's crucial to employ equations of state not contingent on these features.

In our work, we constructed gravity-independent equations of state based on theoretical first principles. Unlike prior research \cite{LopeOter:2019pcq}, where the slope was arbitrarily chosen, we followed the growth rate from the chiral region to determine the slope in the intermediate region. The speed of sound increased progressively until a specific energy density, after which it remained constant before entering the perturbative QCD region. For equations with phase transitions, the slope at the phase transition's end was slightly larger than at the start.

No observational constraints were applied in generating these equations of state, as they often rely on GR-based measurements. Instead, we constrained the range of the parameter $\alpha$ using pulsar measurements. Additionally, we refrained from using other phenomenological model-dependent equations of state due to a lack of uncertainty bands. Our aim was to create equations of state as model-independent as possible, independent of astrophysical observations or GR.

After summarizing Palatini $f(\hat R)$ gravity's key characteristics, we addressed potential issues related to singular parameter values, utilizing these singularities to constrain the parameter within finite ranges for each equation of state.

Notably, we found that positive $\alpha$ values reduced neutron star mass and radius compared to GR, while negative values increased them. However, the range of allowable $\alpha$ values varied depending on the equation of state. Stiffer equations required positive $\alpha$ values, resulting in smoother mass-radius diagrams with a distinct behavior from GR. Softer equations necessitated negative $\alpha$ values for consistency with observations, leading to increased mass and radius.

In summary, for the stiffest equations, $\alpha$ ranged from $275$ to $1030$ km$^2$. For equations within mass bounds, $\alpha$ ranged from $-200$ to $80$ km$^2$. For softer equations, $\alpha$ was constrained from $-325$ to $-200$ km$^2$. For the softest equations, $\alpha$ values varied from $-1500$ to $-880$ km$^2$, albeit only for specific conditions.
Considering these constraints, the parameter $\alpha$ in the Palatini framework falls within the range of  $-325 \lesssim \alpha \lesssim 100$ km$^2$, yielding a constrained $\beta$ of $-6.47 \lesssim \beta \lesssim 1.99$ km$^2$.

We also observed that Palatini gravity can alter mass-radius curves compared to GR, potentially indicating phase transitions. This warrants further investigation.

Furthermore, Palatini gravity allows for soft equations of state \cite{Olmo:2019flu}, previously excluded by GR-based observations. These gravity theories can weaken gravitational attraction depending on their parameter values \cite{Dobado:2011gd}.

Finally, it's worth noting that Palatini gravity exhibits similar phase transition effects in early cosmology \cite{Borowiec:2015qrp,Szydlowski:2015fcq,Stachowski:2016zio}, leading to a "natural" inflationary mechanism.

\acknowledgments

The authors would like to express their sincere gratitude to Felipe Llanes-Estrada for his invaluable insights and enlightening discussions.

The authors acknowledge financial support from MICINN (Spain) {\it Ayuda Juan de la Cierva - incorporaci\'on} 2020 No. IJC2020-044751-I, PID2019-108655GB-I00, PID2019-106080GB-C21 (Spain); COST action CA16214 (Multimessenger Physics and Astrophysics of Neutron Stars); Univ. Complutense de Madrid under research group 910309 and IPARCOS.

This preprint is issued with number IPARCOS-UCM-23-058.

 \bibliographystyle{JHEP}
 \bibliography{biblio.bib}
\end{document}